\documentclass[twocolumn,showpacs]{revtex4}
\pdfoutput=1
\usepackage{graphicx}
\usepackage{dcolumn}
\usepackage{longtable}
\usepackage{amsmath}
\usepackage{amssymb}

\begin{document}

\title
{Octupole deformation in light actinides within an analytic quadrupole octupole axially symmetric model with Davidson potential}

\author
{Dennis Bonatsos$^1$, Andriana Martinou$^1$, N. Minkov$^2$, S. Karampagia$^1$, and D. Petrellis$^{1,3}$}

\affiliation
{$^1$Institute of Nuclear and Particle Physics, National Centre for Scientific Research 
``Demokritos'', GR-15310 Aghia Paraskevi, Attiki, Greece}

\affiliation
{$^2$Institute of Nuclear Research and Nuclear Energy, Bulgarian Academy of Sciences, 72 Tzarigrad Road, 1784 Sofia, Bulgaria}

\affiliation 
{$^3$ Department of Physics, University of Istanbul, 34134 Vezneciler, Istanbul, Turkey}

\begin{abstract}

The analytic quadrupole octupole axially symmetric model, which had successfully predicted $^{226}$Ra and $^{226}$Th as lying at the border between the regions of octupole deformation and octupole vibrations in the light actinides using an infinite well potential (AQOA-IW), 
is made applicable to a wider region of nuclei exhibiting octupole deformation, through the use of a Davidson potential, $\beta^2 + \beta_0^4/\beta^2$
(AQOA-D). Analytic expressions for energy spectra and B(E1), B(E2), B(E3) transition rates are derived. The spectra of $^{222-226}$Ra and 
$^{224,226}$Th are described in terms of the two parameters $\phi_0$ (expressing the relative amount of octupole vs. quadrupole 
deformation) and $\beta_0$ (the position of the minimum of the Davidson potential), while the recently determined B(EL) transition rates
of $^{224}$Ra, presenting stable octupole deformation, are successfully reproduced. A procedure for gradually determining 
the parameters appearing in the B(EL) transitions from a minimum set of data,  thus increasing the predictive power of the model, is outlined.

\end{abstract}



\maketitle
    
\section{Introduction}

Rotational nuclear spectra have long been attributed to quadrupole 
deformations \cite{BM}. However, octupole deformations [corresponding to 
reflection asymmetric (pearlike) shapes] \cite{Rohoz,AB,BN} are supposed to occur in certain regions, 
most notably in the light actinides \cite{Schueler,Wieden,Cocks1,Cocks2} and in some light rare earths \cite{Phillips,Phillips2,Cottle}. 
The hallmark of octupole deformation is
a negative parity band with levels $L^{\pi}=1^-$, $3^-$, $5^-$, \dots, lying 
close to the ground state band and forming with it a single band with 
$L^{\pi} = 0^+$, $1^-$, $2^+$, $3^-$, $4^+$, \dots, while a negative parity 
band lying systematically higher than the ground state band is a footprint 
of octupole vibrations. 

The transition from the regime of octupole vibrations 
into the region of octupole deformation has been considered by several 
authors \cite{Nazar,Nazar2,Sheline}. In the analytic quadrupole octupole axially 
symmetric (AQOA) model \cite{AQOA}, the actinides lying on the border between the regions 
of octupole deformation and octupole vibrations have been described, making the following 
assumptions.

1) Quadrupole and octupole deformations are taken into account on equal 
footing, their relative presence described by the only free parameter in 
the model, $\phi_0$.  

2) Axial symmetry is assumed, in order to keep the problem tractable. 

3) Separation of variables is achieved in a way analogous to the one used
in the framework of the X(5) model \cite{IacX5}, describing the first order shape phase 
transition between spherical and quadrupole deformed shapes \cite{RMP82}.  

4) An infinite well potential is assumed appropriate for the description 
of the border region, as in the E(5) \cite{IacE5} and X(5) \cite{IacX5}
models, the former one describing the second order shape phase transition between 
spherical and $\gamma$-unstable nuclei. Therefore we are going to call this solution
the AQOA-IW model.  

A different approach to the problem of phase transition in the octupole mode
has been developed by Bizzeti and Bizzeti-Sona \cite{Bizzeti70,Bizzeti77}, characterized by
the introduction of a new parametrization of the quadrupole and octupole degrees of freedom, using 
as intrinsic frame of reference the principal axes of the overall tensor 
of inertia, as resulting from the combined quadrupole and octupole deformation.
The main differences between the two models are:

1) The AQOA model is analytic, while the model of Refs. \cite{Bizzeti70,Bizzeti77} is not.

2) In the AQOA model the symmetry axes of the quadrupole and octupole 
deformations are taken to coincide, in order to guarantee axial symmetry, 
while in the more general framework of Refs. \cite{Bizzeti70,Bizzeti77} nonaxial 
contributions, small but not frozen to zero, are taken into account. 

In both models \cite{AQOA,Bizzeti70,Bizzeti77}, $^{226}$Ra and $^{226}$Th appear to lie close to the point of
transition between octupole deformation and octupole vibrations, with heavier isotopes 
corresponding to octupole vibrations and lighter isotopes exhibiting octupole deformation. 

The recent experimental verification of stable octupole deformation in $^{224}$Ra \cite{224Ra} stirred
interest in octupole deformation in the light actinides and their theoretical interpretation.
The AQOA model can be made applicable to deformed nuclei near the transition point 
by replacing the infinite well potential by the Davidson potential \cite{Dav}  of the form 
$\beta^2 +\beta_0^4/\beta^2$,
which contains an additional free parameter, the position $\beta_0$ of the minimum of the potential well.
The flexibility acquired through the replacement of the infinite well potential by the Davidson 
potential has been demonstrated and exploited in the case of quadrupole deformation in \cite{ESD}. 
The analytic quadrupole octupole axially symmetric model with a Davidson potential, to be called the
AQOA-D model, is the subject of the present work. In addition to the spectra of $^{222-226}$Ra \cite{Cocks1,Cocks2} 
and $^{224,226}$Th \cite{Th224,Th226}, the recently measured \cite{224Ra}
electric transition probabilities   of $^{224}$Ra provide an excellent test ground for the model,
already exploited in the Bizzeti and Bizzeti-Sona approach \cite{Bizzeti88}.  

The above mentioned work on the octupole degree of freedom has been developed 
in the framework of the collective model \cite{BM}. Alternative approaches include the following.

1) A complete algebraic classification of the states occurring 
in the simultaneous presence of the quadrupole and octupole degrees of freedom 
has been provided in terms of the spdf-interacting boson model 
\cite{EIPRL,EINPA,Kusnezov1,Kusnezov2}, which has been successfully applied to Ra \cite{Zamfir1}, Th \cite{Zamfir1},
U \cite{Zamfir2}, and Pu \cite{Zamfir2} isotopes. Mean field studies of the critical point for the onset of octupole deformation 
in quadrupole deformed systems have been carried out in Refs. \cite{Kuyucak,Honma}.  

2) An alternative interpretation of the low-lying negative parity states in the light actinides 
has been provided in terms of clustering \cite{Daley,Buck,Buck2,Shneidman,Shneidman2}. The recent experimental findings
for $^{224}$Ra \cite{224Ra} seem to point against this interpretation, but wider evidence in more nuclei is desirable.

3) Relativistic mean field calculations involving the octupole degree of freedom have been carried out both in the light actinides region 
\cite{Geng,Guo} and in the light rare earths \cite{Zhang,Zhang2}, corroborating \cite{Guo,Nomura} the transition from octupole deformation
to octupole vibrations in the light Th isotopes. 

4) A hybrid approach combining the algebraic approach of the interacting boson model of 1) with the relativistic energy density functional theory of 3) 
has been recently developed \cite{Otsuka} and applied the Ra and Th isotopes \cite{Nomura,Nomura2}, as well as to the rare earths Ba and Sm \cite{Nomura2}, 
again corroborating the transition from octupole deformation to octupole vibrations in the light Ra and Th isotopes. 

5) The extended coherent state model (ECSM) has been successfully applied to the description of several negative parity bands 
in Rn \cite{Raduta57}, Ra \cite{Raduta57,Raduta55,Raduta720,Raduta67}, Th \cite{Raduta67}, U \cite{Raduta67}, and Pu \cite{Raduta67} isotopes. 

The AQOA-D model is described in Section II, while in Section III numerical results are provided
and subsequently discussed in Section IV. The integrals needed in the calculation of electric transition 
probabilities are calculated in Appendices A-C, while in Appendix D some details of the derivation of the Hamiltonian,
the method of solution and the comparison to other approaches are given.

\section{The Analytic Quadrupole Octupole Axially Symmetric (AQOA) Model}

\subsection{Formulation} 

In the AQOA model \cite{AQOA} the following assumptions are made: 

a) The axes of the quadrupole and octupole deformations are taken to coincide.
In other words, axial symmetry is assumed, while the $\gamma$ degree of freedom is ignored. 

b) Levels with $K \neq 0$ (where $K$ is the projection of the angular momentum on the 
body-fixed $z'$ axis) are ignored, since they are lying infinitely high in energy \cite{Dzy}.

The Hamiltonian of the AQOA model reads \cite{Dzy,Den}
\begin{multline}\label{eq:e1}
H = -\sum_{\lambda=2,3} {\hbar^2 \over 2 B_\lambda} {1\over \beta_\lambda^3} 
{\partial \over \partial \beta_\lambda} \beta_\lambda^3 {\partial \over 
\partial \beta_\lambda} + {\hbar^2 \hat {L^2} \over 6(B_2 \beta_2^2 + 
2 B_3 \beta_3^2) } \\ + V(\beta_2,\beta_3)  
\end{multline}
where $\beta_2$ and $\beta_3$ are the quadrupole and octupole deformations, $B_2$, $B_3$ are the mass parameters, 
and $\hat L$ is the angular momentum operator in the intrinsic frame, taken along the principal axes of inertia. 	 

The solutions of the Schr\"odinger equation read \cite{Dzy}
\begin{equation}\label{eq:e2}
\Phi^{\pm}_L(\beta_2,\beta_3,\theta) = (\beta_2 \beta_3)^{-3/2} 
\Psi^{\pm}_L(\beta_2,\beta_3) \vert LM0,\pm\rangle,
\end{equation} 
where $\theta$ are the Euler angles describing the orientation of the body-fixed axes 
$x'$, $y'$, $z'$ relative to the laboratory-fixed axes $x$, $y$, $z$, while 
the function $\vert LM0, \pm\rangle$ describes the rotation of an 
axially symmetric nucleus with angular momentum projection $M$ onto the 
laboratory-fixed $z$-axis and projection $K=0$ onto the body-fixed 
$z'$-axis \cite{BM}
\begin{equation}\label{eq:e3}
\vert LM0, \pm\rangle = \sqrt{ 2L+1 \over 32 \pi^2} (1 \pm (-1)^L ) {\cal D}^L_{0,M}(\theta), 
\end{equation} 
with ${\cal D}(\theta)$ denoting Wigner functions of the Euler angles.

Wave functions with the $+$ label correspond to positive parity states 
with $L=0$, 2, 4, \dots, while these with the $-$ label correspond to negative parity states 
with $L=1$, 3, 5, \dots. 

The Schr\"odinger equation can be simplified by introducing \cite{Dzy,Den} 
\begin{equation}\label{eq:e5}
\tilde \beta_2 = \beta_2 \sqrt{B_2\over B}, \quad \tilde \beta_3 = \beta_3 
\sqrt{B_3\over B}, \quad B= {B_2+B_3 \over 2}, 
\end{equation}
 reduced energies $\epsilon=(2B/\hbar^2) E$ and reduced potentials 
$v=(2B/\hbar^2) V$, as well as polar coordinates 
(with $0\leq \tilde \beta < \infty$ and $-\pi/2 \leq \phi \leq \pi/2$) 
\cite{Dzy,Den} 
\begin{equation}\label{eq:e7} 
\tilde \beta_2 = \tilde \beta \cos \phi, \quad \tilde \beta_3 = \tilde \beta
\sin \phi, \quad \tilde \beta = \sqrt{\tilde \beta_2^2 + \tilde \beta_3^2}, 
\end{equation} 
leading to 
\begin{multline}
\left[ -{\partial^2 \over \partial \tilde \beta^2} -{1\over \tilde \beta} 
{\partial \over \partial \tilde \beta} +{L(L+1) \over 3 \tilde \beta^2 
(1+\sin^2\phi) } -{1\over \tilde \beta^2} {\partial^2 \over \partial \phi^2} \right. \\ 
\left. + v(\tilde \beta,\phi) + {3\over  \tilde \beta^2 \sin^2 2\phi}-\epsilon_L
\right] \Psi_L^{\pm}(\tilde \beta,\phi) =0. \label{eq:e8}
\end{multline}

In addition, separation of variables can be achieved by assuming 
the potential to be of the form 
$v(\tilde \beta,\phi) = u(\tilde\beta) + w(\tilde \phi^\pm)$, 
where $w(\tilde \phi^\pm )$ 
is supposed to be of the form of two very steep harmonic oscillators centered at 
the values $\pm \phi_0$. 
In this way Eq. (\ref{eq:e8}) is separated into 
\begin{multline}\label{eq:e9} 
\left[ -{\partial^2 \over \partial \tilde \beta^2} - {1\over \tilde \beta} 
{\partial \over \partial \tilde \beta} +{1\over \tilde\beta^2} \left(
 {L(L+1) \over 3 (1+\sin^2\phi_0)} + {3\over \sin^2 2\phi_0}\right) \right. \\ 
\left. +u(\tilde \beta) -\epsilon_{\tilde\beta}(L) \right] \psi_L (\tilde \beta) 
=0, 
\end{multline}
and 
\begin{equation}\label{eq:e10}
\left[ -{1\over \langle \tilde \beta^2 \rangle}{\partial^2 \over 
\partial(\tilde \phi^\pm)^2} + w (\tilde \phi^\pm ) - \epsilon_{\phi} \right]
 \chi(\tilde \phi^\pm ) =0, 
\end{equation}
where $\Psi_L^{\pm}(\tilde \beta,\phi) = N_{\tilde \beta}\psi_L(\tilde \beta)
N_\phi (\chi(\tilde \phi^+) \pm \chi(\tilde \phi^-))/\sqrt{2}$,
with $N_{\tilde\beta}$ and $N_\phi$ being normalization factors, 
while $\langle \tilde \beta^2\rangle$ is the average of 
$\tilde \beta^2$ over $\psi(\tilde \beta)$, and $\epsilon_L= 
\epsilon_{\tilde \beta}(L) +\epsilon_{\phi}$. 

On the above the following comments apply:

a) $\phi=0$ corresponds to quadrupole deformation alone, 
while $\phi=\pm \pi/2$ corresponds to octupole deformation alone. 

b) Because of the two steep oscillators involved, $\phi$ remains close to $\pm \phi_0$ 
and, therefore, the relative amount of quadrupole and octupole deformation remains constant. 

Some details of the derivation of the Hamiltonian, the method of solution, and the comparison of the present model
to other approaches are given in Appendices D1-D4.

\subsection{The $\tilde \beta$ part of the spectrum}

Eq. (\ref{eq:e9}) is exactly soluble \cite{Elliott,Rowe} in the case of 
the Davidson potentials
 \cite{Dav} 
\begin{equation}\label{eq:e15} 
u(\tilde \beta) = \tilde \beta^2 +{\tilde \beta_0^4 \over \tilde \beta^2},
\end{equation}
in which the eigenfunctions are Laguerre polynomials 
\begin{equation}\label{eq:e16}
F^L_{n_\beta}(\tilde \beta) = \sqrt{2n_\beta! \over \Gamma( n_\beta+a_L+1)} \tilde\beta^{a_L} 
L_{n_\beta}^{a_L}(\tilde \beta^2) e^{-\tilde \beta^2/2}, 
\end{equation}
with
\begin{equation}\label{eq:e17}
a_L= \sqrt{ {L(L+1)\over 3(1+\sin^2\phi_0)} + {3\over \sin^2 2\phi_0} +\beta_0^4}
\end{equation}
while the energy eigenvalues are given by 
\begin{multline}\label{eq:e18}
E_{n_\beta,L} = 2n_\beta+a_L +1 = 2n_\beta+1 \\
+ \sqrt{ {L(L+1)\over 3(1+\sin^2 \phi_0)} + {3\over 
\sin^2 2\phi_0} +\beta_0^4} . 
\end{multline}

Eq. (\ref{eq:e9}) is also exactly soluble in the case of an infinite well potential in $\tilde \beta$,
in which the eigenfunctions are Bessel functions. This solution has been worked out in Ref. \cite{AQOA}. 

\subsection{The $\phi$ part of the spectrum}

Eq. (\ref{eq:e10}) with the potential 
corresponding to two harmonic oscillators centered at $\pm \phi_0$ 
\begin{equation}\label{eq:e8a}
w (\tilde \phi^\pm )= {1\over 2} c (\phi \mp \phi_0)^2 = {1\over 2} c 
 (\tilde \phi^\pm)^2, \qquad \tilde \phi^\pm = \phi \mp \phi_0,  
\end{equation}
has been solved in Ref. \cite{AQOA}.
The energy eigenvalues are
\begin{equation}\label{eq:e21}
\epsilon_{\phi} = \sqrt{ 2c\over \langle \tilde \beta^2\rangle } \left(
n_\phi +{1\over 2} \right), \qquad n_\phi = 0,1,2,\ldots 
\end{equation}
where $n_\phi$ is the number of quanta in the $\phi$ degree of freedom, while 
the eigenfunctions are Hermite polynomials $H_{n_\phi}$
\begin{multline}\label{eq:e22}
\chi_{n_\phi}(\tilde \phi^\pm) = N_{n_\phi} H_{n_\phi} (b \tilde \phi^\pm) 
e^{-b^2 (\tilde \phi^\pm)^2 /2} , \\ b=\left( c \langle \tilde \beta^2 
\rangle \over 2 \right)^{1/4},
\end{multline}
with normalization constant $ N_{n_\phi} = \sqrt{b\over \sqrt{\pi} 2^{n_\phi}
n_\phi!}$.

The total energy in the present model is then 
\begin{multline}\label{eq:e24}
E(n_\beta,L,\phi_0, n_\phi) = E_0 + C_1 E_{n_\beta,L} + C_2 n_\phi,\\
C_1= {\hbar^2 \over 2B}, \qquad C_2= {\hbar^2 \over 2B} \sqrt{{2c\over \langle \beta^2 \rangle}}.
\end{multline}
In what follows only bands with $n_\phi=0$ will be considered. 

\begin{figure}
\includegraphics[width=0.47\textwidth]{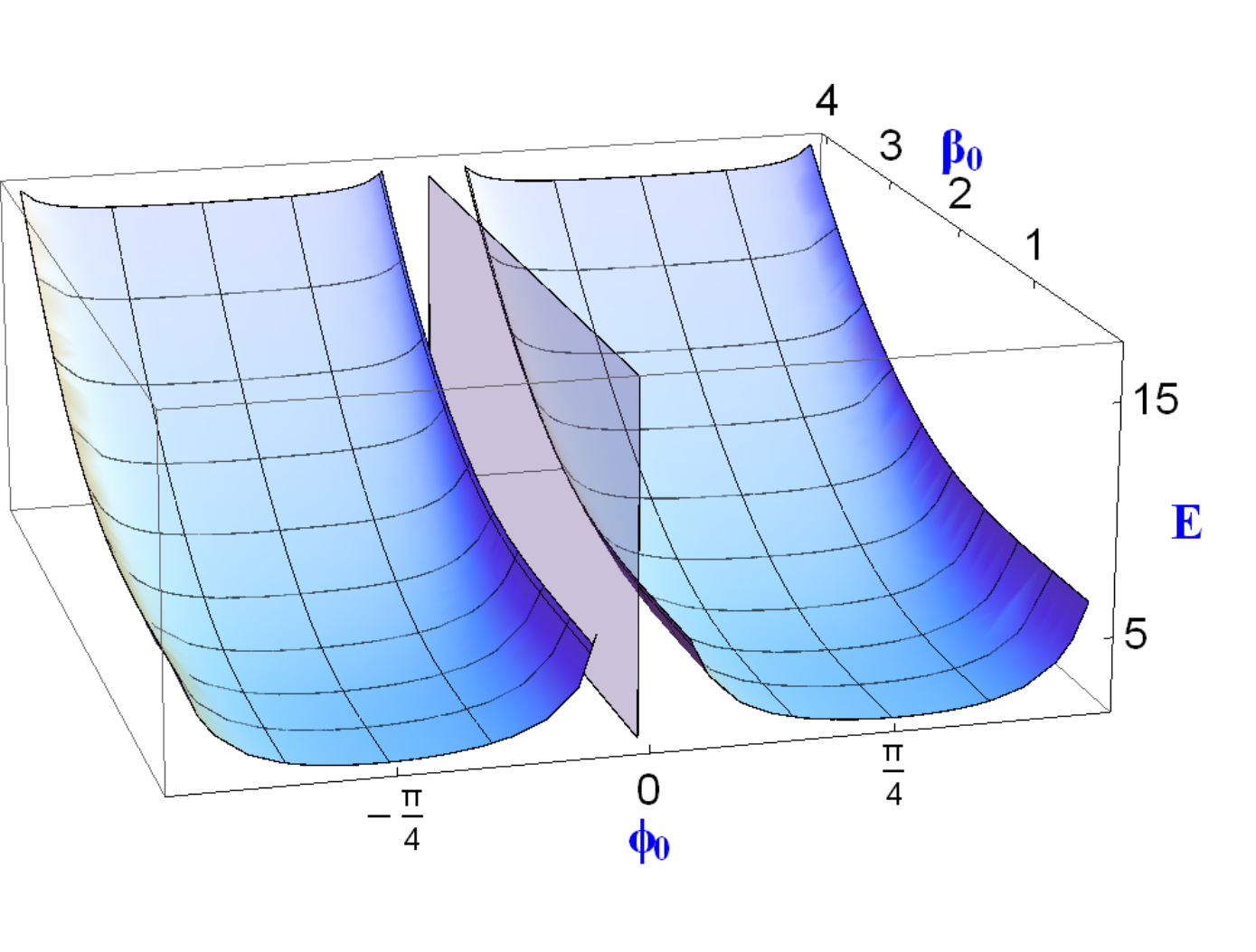}\vspace{0.2cm}
\centering \caption{(Color online) Excitation energies $E_{n_\beta,L}$ [Eq. (\ref{eq:e18})] for $n_\beta=0$ 
and $L=2$ are shown as a function of the free parameters $\beta_0$ and $\phi_0$. All quantities shown 
are dimensionless. See subsection II.C for further discussion.
}
\end{figure}

As an example of the dependence of the energy levels on the free parameters $\beta_0$ and $\phi_0$ in Eq. (\ref{eq:e18}), 
the energy levels with $n_{\beta}=0$ and $L=2$ are shown in Fig. 1. 
Smooth variation with both parameters is seen. It is worth 
remarking that very slight dependence on $\phi_0$ is observed between $\pi/8$ and $3\pi/8$, in agreement 
with the findings of Ref. \cite{AQOA}. This observation (partly) justifies {\it a posteriori} 
the adiabatic approximation used in relation to the $\phi$ degree of freedom, described in Appendix D3. 

\subsection{$B(EL)$ transition rates}

The electric quadrupole and octupole operators are (Eq. (6-63) of \cite{BM}) 
\begin{multline}\label{eq:e31}
T^{(E2)}_\mu = t_2 \beta_2 {\cal D}^{(2)}_{0,\mu}(\theta)= t_2 \sqrt{B\over B_2} \tilde \beta \cos\phi {\cal D}^{(2)}_{0,\mu}(\theta), \\ 
T^{(E3)}_\mu =t_3 \beta_3 {\cal D}^{(3)}_{0,\mu}(\theta)= t_3 \sqrt{B\over B_3} \tilde \beta \sin\phi {\cal D}^{(3)}_{0,\mu}(\theta), 
\end{multline}
with
\begin{equation}\label{t2t3}
t_2= {3 Z e\over 4\pi} R^2, \qquad t_3= {3 Z e\over 4\pi} R^3,
\end{equation}
where $R$ is the effective radius of the nucleus, 
while the electric dipole operator reads \cite{Dzy}
\begin{equation}\label{eq:e32}
T^{(E1)}_\mu = t_1 \beta_2 \beta_3 {\cal D}^{(1)}_{0,\mu}(\theta) = t_1 {B\over \sqrt{B_2 
B_3}} \tilde\beta^2 {\sin 2\phi \over 2} {\cal D}^{(1)}_{0,\mu}(\theta).
\end{equation}

The total wave function reads 
\begin{multline}\label{eq:e33}
\Phi^{\pm}_{L,M,q} (\tilde\beta, \phi, \theta) = (\beta_2 \beta_3)^{-3/2} 
N_{\tilde \beta} F_{n_\beta}^L(\tilde \beta) N_\phi \\
{(\chi_{n_\phi}(\tilde \phi^+)\pm \chi_{n_\phi}(\tilde \phi^-))\over \sqrt{2}} 
\sqrt{2L+1\over 32 \pi^2} (1\pm (-1)^L) {\cal D}^L_{0,M}(\theta),
\end{multline} 
where $q$ stands for the rest of the quantum numbers ($n_\beta$, $n_\phi$). 
Since in what follows only bands with $n_\phi=0$ will be considered, in the remainder of the paper, as well 
as in the Appendices, we simplify the notation by using  $n$ instead of $n_\beta$. As a result, in what follows,
$n_i$ and $n_f$ indicate the initial and final values of $n_\beta$, while $q_i$ and $q_f$ denote $n_i$ and $n_f$ 
respectively, with $n_\phi=0$  at all times. 

$B(EL)$ transition rates are given by 
\begin{equation}\label{eq:e34} 
B(EL; L_i q_i \to L_f q_f) = { \vert \langle L_f q_f \vert \vert T^{(EL)}
\vert \vert L_i q_i \rangle \vert^2 \over (2L_i+1)}, 
\end{equation}
where the reduced matrix element is obtained through the Wigner-Eckart theorem
\begin{multline}\label{eq:e35} 
\langle L_f \mu_f q_f \vert T^{(EL)}_\mu \vert L_i \mu_i q_i\rangle = \\
{ (L_i L L_f \vert \mu_i \mu \mu_f)\over \sqrt{2L_f+1} } \langle L_f q_f 
\vert \vert T^{EL} \vert \vert L_i q_i \rangle . 
\end{multline}

In Eq. (\ref{eq:e34}) the integration over the Euler angles $\theta$ 
involves standard integrals over three Wigner functions calculated in Appendix B, 
while the rest of the integrations are performed over 
$\int\int \beta_2^3 d\beta_2 \beta_3^3 d\beta_3$, where the $\beta_2^3$, 
$\beta_3^3$ factors come from the volume element and cancel with the 
first factor of Eq. (\ref{eq:e33}). Using Eqs. (\ref{eq:e5}) and 
(\ref{eq:e7}), as well as the relevant Jacobian, one finds (up to constant 
factors) that the integration is over $\int\int d\beta_2 d\beta_3 = {B\over \sqrt{B_2 B_3}} \int\int \tilde \beta d\tilde \beta d\phi$. 

Relevant integrals over $\phi$ are calculated in Appendix A, while integrals over $\tilde \beta$ are determined in Appendix C. 
The final results for matrix elements are summarized here.

Matrix elements of $T^{(E2)}$ between positive parity levels within the ground state band ($n_i=n_f=0$) read 
\begin{multline}\label{MESS}
\langle L_f q_f \vert \vert T^{E2} \vert \vert L_i q_i \rangle_{S\to S} = 
t_2 \sqrt{B\over B_2} e^{-{1\over 4b^2}} { \cos\phi_0+ e^{-b^2 \phi_0^2} \over 1+ e^{-b^2 \phi_0^2}} \\
\sqrt{2L_i+1}(L_i 2 L_f | 0 0 0)
\frac{\Gamma\left(\frac{a_{i}}{2}+\frac{a_{f}}{2}+\frac{3}{2}\right)}
{\sqrt{\Gamma(a_{i}+1)\Gamma(a_{f}+1)}}.
\end{multline}

Matrix elements of $T^{(E2)}$ from a positive parity level of the first excited band ($n_i=1$) 
to a positive parity level of the ground state band ($n_f=0$) are
\begin{multline}\label{MESS10}
\langle L_f q_f \vert \vert T^{E2} \vert \vert L_i q_i \rangle_{S\to S} = 
t_2 \sqrt{B\over B_2} \\ e^{-{1\over 4b^2}} { \cos\phi_0+ e^{-b^2 \phi_0^2} \over 1+ e^{-b^2 \phi_0^2}}
\sqrt{2L_i+1}(L_i 2 L_f | 0 0 0) \\
\frac{\Gamma\left(\frac{a_{i}}{2}+\frac{a_{f}}{2}+\frac{3}{2}\right) \Gamma\left(\frac{a_{i}}{2}-\frac{a_{f}}{2}+\frac{1}{2}\right)}
{\sqrt{\Gamma(a_{i}+2)\Gamma(a_{f}+1)} \Gamma\left(\frac{a_{i}}{2}-\frac{a_{f}}{2}-\frac{1}{2}\right)}.
\end{multline}

Matrix elements of $T^{(E2)}$ between negative parity levels within the lowest band ($n_i=n_f=0$) have the form 
\begin{multline}\label{MEAA}
\langle L_f q_f \vert \vert T^{E2} \vert \vert L_i q_i \rangle_{A\to A} = 
 t_2 \sqrt{B\over B_2} e^{-{1\over 4b^2}} { \cos\phi_0- e^{-b^2 \phi_0^2} \over 1- e^{-b^2 \phi_0^2}} \\
\sqrt{2L_i+1} (L_i 2 L_f | 0 0 0)
\frac{\Gamma\left(\frac{a_{i}}{2}+\frac{a_{f}}{2}+\frac{3}{2}\right)}
{\sqrt{\Gamma(a_{i}+1)\Gamma(a_{f}+1)}}.
\end{multline}

Matrix elements of $T^{(E3)}$ between a positive parity level of the ground state band and a negative parity level of the lowest band,
or vice versa, read
\begin{multline}\label{MEE3}
\langle L_f q_f \vert \vert T^{E3} \vert \vert L_i q_i \rangle = 
 t_3 \sqrt{B\over B_3} {e^{-{1\over 4 b^2}} \sin\phi_0 \over \sqrt{1- e^{-2b^2 \phi_0^2}}}
\sqrt{2L_i+1}\\ (L_i 3 L_f | 0 0 0)
\frac{\Gamma\left(\frac{a_{i}}{2}+\frac{a_{f}}{2}+\frac{3}{2}\right)}
{\sqrt{\Gamma(a_{i}+1)\Gamma(a_{f}+1)}}.
\end{multline}

Matrix elements of $T^{(E1)}$ between a positive parity level of the ground state band and a negative parity level of the lowest band,
or vice versa, are 
\begin{multline}\label{MEE1}
\langle L_f q_f \vert \vert T^{E1} \vert \vert L_i q_i \rangle_{S\to S} = 
{1\over 2} t_1 {B \over \sqrt{B_2 B_3}} {e^{-{1\over  b^2}} \sin 2\phi_0 \over \sqrt{1- e^{-2b^2 \phi_0^2}}} \\
\sqrt{2L_i+1}(L_i 1 L_f | 0 0 0)
\frac{\Gamma\left(\frac{a_{i}}{2}+\frac{a_{f}}{2}+2\right)}
{\sqrt{\Gamma(a_{i}+1)\Gamma(a_{f}+1)}}.
\end{multline}

The final results for $B(EL)$s are summarized here.

$B(E2)$s between positive parity levels within the ground state band ($n_i=n_f=0$) read 
\begin{multline}\label{BE2SS}
B(E2; L_i q_i\to L_f q_f)_{S\to S}= t_2^2 {B\over B_2} \\ e^{-{1\over 2b^2}} { (\cos\phi_0+ e^{-b^2 \phi_0^2})^2 \over (1+ e^{-b^2 \phi_0^2})^2} \\
(L_i 2 L_f | 0 0 0)^2
\frac{\left[\Gamma\left(\frac{a_{i}}{2}+\frac{a_{f}}{2}+\frac{3}{2}\right)\right]^2}
{\Gamma(a_{i}+1)\Gamma(a_{f}+1)}.
\end{multline}

$B(E2)$s from a positive parity level of the first excited band ($n_i=1$) 
to a positive parity level of the ground state band ($n_f=0$) are 
\begin{multline}\label{BE2SS10}
B(E2; L_i q_i\to L_f q_f)_{S\to S}= t_2^2 {B\over B_2} \\
e^{-{1\over 2b^2}} { (\cos\phi_0+ e^{-b^2 \phi_0^2})^2 \over (1+ e^{-b^2 \phi_0^2})^2}
(L_i 2 L_f | 0 0 0)^2 \\
\frac{\left[\Gamma\left(\frac{a_{i}}{2}+\frac{a_{f}}{2}+\frac{3}{2}\right) 
\Gamma\left(\frac{a_{i}}{2}-\frac{a_{f}}{2}+\frac{1}{2}\right)\right]^2}
{\Gamma(a_{i}+2)\Gamma(a_{f}+1) \left[\Gamma\left(\frac{a_{i}}{2}-\frac{a_{f}}{2}-\frac{1}{2}\right)\right]^2 }.
\end{multline}

$B(E2)$s between negative parity levels within the lowest band ($n_i=n_f=0$) have the form 
\begin{multline}\label{BE2AA}
B(E2; L_i q_i\to L_f q_f)_{A\to A}= t_2^2 {B\over B_2} \\ e^{-{1\over 2b^2}} { (\cos\phi_0- e^{-b^2 \phi_0^2})^2 \over (1- e^{-b^2 \phi_0^2})^2} \\
(L_i 2 L_f | 0 0 0)^2
\frac{\left[\Gamma\left(\frac{a_{i}}{2}+\frac{a_{f}}{2}+\frac{3}{2}\right)\right]^2}
{\Gamma(a_{i}+1)\Gamma(a_{f}+1)}.
\end{multline}

$B(E3)$s between a positive parity level of the ground state band and a negative parity level of the lowest band,
or vice versa, read 
\begin{multline}\label{BE3}
B(E3; L_i q_i\to L_f q_f)= t_3^2 {B\over B_3} {e^{-{1\over 2 b^2}} \sin^2\phi_0 \over (1- e^{-2b^2 \phi_0^2})} \\
(L_i 3 L_f | 0 0 0)^2
\frac{\left[\Gamma\left(\frac{a_{i}}{2}+\frac{a_{f}}{2}+\frac{3}{2}\right)\right]^2}
{\Gamma(a_{i}+1)\Gamma(a_{f}+1)}.
\end{multline}

$B(E1)$s between a positive parity level of the ground state band and a negative parity level of the lowest band,
or vice versa, are 
\begin{multline}\label{BE1}
B(E1; L_i q_i\to L_f q_f)={1\over 4} t_1^2 {B^2 \over B_2 B_3} {e^{-{2\over  b^2}} \sin^2 2\phi_0 \over (1- e^{-2b^2 \phi_0^2})} \\
(L_i 1 L_f | 0 0 0)^2 \frac{\left[\Gamma\left(\frac{a_{i}}{2}+\frac{a_{f}}{2}+2\right)\right]^2}
{\Gamma(a_{i}+1)\Gamma(a_{f}+1)}.
\end{multline}

In Ref. \cite{Nomura2} it has been pointed out that $B(E1)$s for transitions from positive parity levels $L$ of the ground 
state band to negative parity levels $L-1$ of the lowest band and $B(E1)$s for transitions in the opposite direction,
i.e. from negative parity levels $L-1$ of the lowest band to positive parity levels $L$ of the ground state band, should be 
of the same order, as seen experimentally. This condition is clearly fulfilled by Eq. (\ref{BE1}). 

\begin{table*}

\caption{Parameters $\phi_0$ and $\beta_0$ of the AQOA model with Davidson potential (AQOA-D) obtained from rms fits 
to experimental spectra of $^{222}$Ra \cite{Cocks1,Cocks2}, $^{224}$Ra \cite{Cocks1,Cocks2}, $^{226}$Ra \cite{Cocks1,Cocks2},
$^{224}$Th \cite{Th224}, and $^{226}$Th \cite{Th226}. 
The experimental $R_{4/2}=E(4_1^+)/E(2_1^+)$ ratios are also shown.  The
angular momenta of the highest levels of the ground state, $\beta$
and negative parity bands included in the rms fit are labelled by $L_g$,
$L_\beta$, and $L_o$ respectively, while $n$ indicates the
total number of levels involved in the fit and $\sigma$ is the
quality measure of Eq. (\ref{qual}). The theoretical predictions are obtained 
from the formulae mentioned in subsec. II.B~.
See subsec. III.A for further discussion. For completeness, the parameter $b$, and the parameter ratios
$t_3/t_2$, and $B_2/B_3$, appearing in the fitting of electric transition rates, are shown, wherever known.
All quantities shown are dimensionless, exept $t_3/t_2$, which is given in fm. 
See subsec. III.B for further discussion.}

\bigskip

\begin{tabular}{ r c r c  c c c c   c c c c  }
\hline nucleus & $R_{4/2}$ & $\phi_0$ & $\beta_0$ & $L_g$ & $L_\beta$ & $L_o$ & $n$ & $\sigma$ & $b$ & $t_3/t_2$(fm) 
& $B_2/B_3$ \\

\hline
$^{222}$Ra & 2.715 & 62.7 & 0.00 & 20 & 0 & 19 & 20 & 0.917 &       & 7.266 &      \\
$^{224}$Ra & 2.970 & 41.9 & 1.80 & 28 & 0 & 27 & 28 & 1.351 & 1.836 & 7.288 & 0.65 \\
$^{226}$Ra & 3.127 & 24.7 & 2.14 & 28 & 0 & 27 & 28 & 1.360 &       & 7.309 &      \\

$^{224}$Th & 2.896 & 67.9 & 0.79 & 18 &   & 17 & 17 & 0.843 &       & 7.288 &      \\
$^{226}$Th & 3.136 &  9.5 & 0.94 & 20 & 0 & 19 & 20 & 0.994 &       & 7.309 &      \\

\hline
\end{tabular}
\end{table*}

\begin{table*}

\caption{Comparison of theoretical predictions of the AQOA model with Davidson potential (AQOA-D, columns labelled as D) 
and with an infinite well potential (AQOA-IW, columns labelled as IW) 
to experimental data [normalized to $E(2_1^+)$] of $^{222}$Ra \cite{Cocks1,Cocks2}, $^{224}$Ra \cite{Cocks1,Cocks2}, $^{226}$Ra \cite{Cocks1,Cocks2},
$^{224}$Th \cite{Th224}, and $^{226}$Th \cite{Th226}. The AQOA-D parameters are shown in Table I, 
while the AQOA-IW predictions have been taken from Ref. \cite{AQOA}. The theoretical predictions for AQOA-D are obtained 
from the formulae mentioned in subsec. II.B~. See subsec. III.A for further discussion. }

\bigskip

\begin{tabular}{ r r r r  r r r r  r r r r r }
\hline 
    & $^{222}$Ra & $^{222}$Ra & $^{224}$Ra & $^{224}$Ra & $^{226}$Ra & $^{226}$Ra & $^{226}$Ra & $^{224}$Th & $^{224}$Th & $^{226}$Th & $^{226}$Th & $^{226}$Th \\
$L^{\pi}$ & exp.       & D       & exp.       & D        & exp.       & D       & IW       & exp.       & D        & exp.       & D        & IW \\

\hline

$ 4^+$ &  2.72 &  3.00 &  2.97 &  3.17 &  3.13 &  3.22 &  3.09 &  2.90 &  3.09 &  3.14 &  3.22 &  3.12 \\ 
$ 6^+$ &  4.95 &  5.59 &  5.68 &  6.21 &  6.16 &  6.45 &  5.99 &  5.45 &  5.90 &  6.20 &  6.44 &  6.10 \\
$ 8^+$ &  7.58 &  8.49 &  8.94 &  9.87 &  9.89 & 10.45 &  9.56 &  8.50 &  9.17 & 10.00 & 10.42 &  9.78 \\
$10^+$ & 10.55 & 11.58 & 12.66 & 13.94 & 14.19 & 15.02 & 13.71 & 11.97 & 12.71 & 14.41 & 14.97 & 14.08 \\
$12^+$ & 13.82 & 14.77 & 16.74 & 18.30 & 18.93 & 20.00 & 18.42 & 15.80 & 16.42 & 19.32 & 19.93 & 18.96 \\
$14^+$ & 17.39 & 18.04 & 21.17 & 22.85 & 24.06 & 25.31 & 23.64 & 19.97 & 20.25 & 24.68 & 25.20 & 24.38 \\
$16^+$ & 21.21 & 21.36 & 25.90 & 27.55 & 29.52 & 30.84 & 29.38 & 24.44 & 24.16 & 30.41 & 30.69 & 30.34 \\
$18^+$ & 25.28 & 24.70 & 30.92 & 32.34 & 35.30 & 36.54 & 35.61 & 29.20 & 28.13 & 36.50 & 36.35 & 36.81 \\
$20^+$ & 29.57 & 28.07 & 36.22 & 37.22 & 41.38 & 42.38 & 42.33 &       &       & 42.90 & 42.14 & 43.80 \\
$22^+$ &       &       & 41.74 & 42.15 & 47.75 & 48.32 & 49.54 &       &       &       &       &       \\
$24^+$ &       &       & 47.48 & 47.13 & 54.44 & 54.34 & 57.22 &       &       &       &       &       \\
$26^+$ &       &       & 53.41 & 52.14 & 61.42 & 60.43 & 65.38 &       &       &       &       &       \\
$28^+$ &       &       & 59.54 & 57.18 & 68.70 & 66.57 & 74.01 &       &       &       &       &       \\

$ 0^+$ &  8.23 &  8.06 & 10.86 & 10.90 & 12.19 & 12.21 & 11.23 &       &  9.91 & 11.18 & 11.31 & 12.41 \\

$ 1^-$ &  2.18 &  0.35 &  2.56 &  0.34 &  3.75 &  0.34 &  0.34 &  2.56 &  0.34 &  3.19 &  0.34 &  0.34 \\
$ 3^-$ &  2.85 &  1.90 &  3.44 &  1.95 &  4.75 &  1.97 &  1.93 &  3.11 &  1.93 &  4.26 &  1.97 &  1.94 \\
$ 5^-$ &  4.26 &  4.24 &  5.13 &  4.60 &  6.60 &  4.72 &  4.45 &  4.74 &  4.43 &  6.24 &  4.72 &  4.51 \\
$ 7^-$ &  6.33 &  7.01 &  7.59 &  7.98 &  9.26 &  8.36 &  7.70 &  7.13 &  7.49 &  9.11 &  8.35 &  7.86 \\
$ 9^-$ &  8.92 & 10.02 & 10.73 & 11.86 & 12.68 & 12.67 & 11.57 & 10.17 & 10.91 & 12.79 & 12.64 & 11.86 \\
$11^-$ & 11.97 & 13.17 & 14.46 & 16.09 & 16.74 & 17.47 & 16.00 & 13.73 & 14.55 & 17.15 & 17.41 & 16.45 \\
$13^-$ & 15.38 & 16.40 & 18.68 & 20.56 & 21.39 & 22.63 & 20.96 & 17.72 & 18.32 & 22.11 & 22.53 & 21.61 \\
$15^-$ & 19.11 & 19.69 & 23.31 & 25.18 & 26.54 & 28.05 & 26.45 & 22.07 & 22.20 & 27.55 & 27.92 & 27.30 \\
$17^-$ & 23.11 & 23.03 & 28.27 & 29.93 & 32.13 & 33.67 & 32.43 & 26.71 & 26.14 & 33.42 & 33.50 & 33.51 \\
$19^-$ & 27.35 & 26.39 & 33.51 & 34.77 & 38.10 & 39.44 & 38.91 &       &       & 39.63 & 39.23 & 40.24 \\
$21^-$ &       &       & 38.99 & 39.68 & 44.41 & 45.34 & 45.87 &       &       &       &       &       \\
$23^-$ &       &       & 44.67 & 44.63 & 51.03 & 51.32 & 53.32 &       &       &       &       &       \\
$25^-$ &       &       & 50.55 & 49.63 & 57.93 & 57.38 & 61.24 &       &       &       &       &       \\
$27^-$ &       &       & 56.60 & 54.66 & 65.08 & 63.49 & 69.64 &       &       &       &       &       \\

\hline
\end{tabular}
\end{table*}

\section{Numerical results}

\subsection{Spectra}

In the spectra of the AQOA model with Davidson potential (AQOA-D) only the parameters $\phi_0$ and $\beta_0$
play an essential role, as seen in Eq. (\ref{eq:e18}), while the quantities $E_0$, $C_1$, and $C_2$ of Eq. (\ref{eq:e24}) 
do not enter, if we consider only bands with $n_\phi=0$ and normalize all energies to that of the first excited state,
$E(2^+_1)$. The parameters of rms fits, using the quality measure 
\begin{equation}\label{qual}
\sigma = \sqrt{ { \sum_{i=1}^n (E_i(exp)-E_i(th))^2 \over (n-1)E(2_1^+)^2 } },
\end{equation}
to the spectra of the Ra and Th isotopes lying at the border between the regions 
of octupole deformation and octupole vibrations, as well as within the former region, are shown in Table I, while in Table II 
the relevant spectra are shown.  For $^{226}$Ra and $^{226}$Th, the predictions of the original one-parameter ($\phi_0$) 
AQOA with an infinite well potential (AQOA-IW), applicable at the border between the regions 
of octupole deformation and octupole vibrations, are shown for comparison.  The following comments apply.   

a) Good agreement between the theoretical predictions of AQOA-D and experimental data is obtained
up to high angular momenta, both in the ground state band and in the negative parity band. 
The predictions for the $1^-$ and $3^-$ states are poor, since no finite barrier is used in the phi potential
\cite{Jolos49,Jolos587,Jolos60}

b) When moving from the border region towards the interior of the region of octupole deformation, the parameter $\phi_0$ increases, in agreement with an increasing role of the octupole deformation (which is proportional to $\sin\phi_0$), while the parameter $\beta_0$ decreases. In parallel, a decreasing role of the quadrupole deformation is revealed by the decreasing $R_{4/2}$ ratios. 
  
c) The agreement between the predictions of AQOA-IW and the data is comparable to that of AQOA-D, 
except at high angular momenta, where AQOA-D is closer to the data, due to the term $\beta_0^4$ appearing 
in Eq. (\ref{eq:e18}), which moderates the increase of the energy with $L$.  

\subsection{Transitions}

With the parameters $\phi_0$ and $\beta_0$ determined from the spectra, we now turn attention 
to electromagnetic transition rates, following the procedure described below. 

a) From Eqs. (\ref{BE2SS}) and (\ref{BE2SS10}) it is clear that ratios of $B(E2)_{S\to S}$s involve only the parameters 
$\phi_0$ and $\beta_0$, thus they are already fixed. The same holds separately for ratios of $B(E2)_{A\to A}$s, 
or $B(E3)$s, or $B(E1)$s, as seen from Eqs. (\ref{BE2AA}), (\ref{BE3}), and (\ref{BE1}) respectively. 

b) Ratios of $B(E2)_{S\to S}$s over $B(E2)_{A\to A}$s involve in addition the parameter $b$, 
which can then be determined from such ratios, as seen from Eqs. (\ref{BE2SS}) and (\ref{BE2AA}). 

c) Ratios of $B(E3)$s over $B(E2)$s involve in addition the ratio $t_3/ t_2$, which can be determined from the 
nuclear radius (see subsec. III.B.2), and the ratio 
$B_2/ B_3$, which can be determined from the $B(E3)/B(E2)$ ratios, as seen from Eqs. (\ref{BE3}) and (\ref{BE2SS}).

This procedure can be tested against the recently measured transition matrix elements of $^{224}$Ra \cite{224Ra},
shown in Table III. 

\begin{table}

\caption{Matrix elements of electric transitions in $^{224}$Ra. The experimental data, in units of $e$ fm, $e$ fm$^2$, $e$ fm$^3$ for 
E1, E2, E3 respectively, have been taken from Ref. \cite{224Ra}, while the theoretical predictions have been obtained using the formulae 
of subsec. II.D in the way described in subsec. III.B, where further discussion is given.}

\bigskip

\begin{tabular}{ l l l  }
\hline m.e. & exp. & th.  \\

\hline

$\langle 0^+ || E2 || 2^+\rangle$        & 199$\pm$3       & 196   \\
$\langle 2^+ || E2 || 4^+\rangle$        & 315$\pm$6       & 323   \\
$\langle 4^+ || E2 || 6^+\rangle$        & 405$\pm$15      & 426   \\
$\langle 6^+ || E2 || 8^+\rangle$        & 500$\pm$60      & 525   \\
         
$\langle 1^- || E2 || 3^-\rangle$        & 230$\pm$11      & 236   \\
$\langle 3^- || E2 || 5^-\rangle$        & 410$\pm$60      & 334   \\

$\langle 0^+ || E2 || 2^+_\gamma\rangle$ & 23$\pm$4        & 36    \\

$\langle 0^+ || E3 || 3^-\rangle$        & 940$\pm$30      & 1006  \\
$\langle 2^+ || E3 || 1^-\rangle$        & 1370$\pm$140    & 1137  \\
$\langle 2^+ || E3 || 3^-\rangle$        & $<$4000         & 1176  \\
$\langle 2^+ || E3 || 5^-\rangle$        & 1410$\pm$190    & 1594  \\ 

$\langle 0^+ || E1 || 1^-\rangle$        & $<$0.018        & 0.013 \\
$\langle 2^+ || E1 || 1^-\rangle$        & $<$0.03         & 0.018 \\
$\langle 2^+ || E1 || 3^-\rangle$        & 0.026$\pm$0.005 & 0.023 \\
$\langle 4^+ || E1 || 5^-\rangle$        & 0.030$\pm$0.010 & 0.032 \\
$\langle 6^+ || E1 || 7^-\rangle$        & $<$0.10         & 0.042 \\

\hline
\end{tabular}
\end{table}

\subsubsection{$E2$ transitions}

The ratio of any $E2_{A\to A}$ matrix element [Eq. (\ref{MEAA})] over any $E2_{S\to S}$ matrix element [Eq. (\ref{MESS})] contains the ratio 
\begin{equation}\label{ra}
r= {(\cos\phi_0 -a)(1+a) \over (\cos\phi_0+a)(1-a)}, \qquad a = e^{-b^2 \phi_0^2}.
\end{equation} 
In the case of $^{224}$Ra, the ratios of experimental matrix elements ${3^-\to 1^-\over 2^+\to 0^+}$, 
${3^-\to 1^-\over 4^+\to 2^+}$, ${3^-\to 1^-\over 6^+\to 4^+}$, ${3^-\to 1^-\over 8^+\to 6^+}$
lead to $r=0.852$, 0.887, 0.909, 0.908, i.e., to an average value of 0.889~.  

Solving for $a$ one obtains the quadratic equation
\begin{equation}
(r-1)a^2 + (r+1)(\cos\phi_0-1)a+(1-r) \cos\phi_0=0,  
\end{equation}
having the solution
\begin{multline}\label{asol}
a= {1\over 2(r-1)}    \left[ -(r+1)(\cos\phi_0-1) \right. \\
\left. \pm \sqrt{ (r+1)^2 (\cos\phi_0-1)^2+4(r-1)^2 \cos\phi_0} \right] . 
\end{multline} 
Since from Eq. (\ref{ra}) one has 
\begin{equation}
b^2 = -{\ln a\over \phi_0^2}, 
\end{equation}
$\ln a$ has to be negative for $b$ to be real. Since $r<1$, we see that in Eq. (\ref{asol}) only the negative sign is allowed
in order to have $a>0$, 
leading in the case of $^{224}$Ra to $a=0.1647$ and $b=1.836$~. 

One can then keep 
\begin{equation}
F=t_2\sqrt{B/B_2} 
\end{equation}
as an overall constant for all $E2$ transition matrix elements,
and determine it through rms fitting to the experimental data of the transitions with $n_i=n_f=0$, obtaining $F=127.20$ 
and the $E2$ predictions reported in Table III. Then from Eq. (\ref{MESS10}) one can calculate also the transitions with $n_i=1$, $n_f=0$, 
one of which is also reported in Table III. 

\subsubsection{$E3$ transitions} 

For the coefficients $t_2$ and $t_3$ one can use Eq. (\ref{t2t3}), leading to 
\begin{equation}
{t_3\over t_2}= R, 
\end{equation}
where $R$ is the nuclear radius, given by \cite{Heyde}
\begin{equation}
R= r_0 A^{1/3}, \qquad r_0=1.2 \ {\rm fm}, 
\end{equation}
with $A$ being the mass number of the nucleus.
Thus in the case of $^{224}$Ra one has $R=7.2878$ fm. 

One can then determine the ratio $B_2\over B_3$ from any ratio of $E3$ matrix element 
over $E2$ matrix element, since each of these ratios contains the quantity 
$(t_3/t_2)\sqrt{B_2/B_3}$, as seen from Eqs. (\ref{MEE3}), (\ref{MESS}), (\ref{MEAA}).  
In the case of $^{224}$Ra, six $E2$ matrix elements and three $E3$ 
matrix elements are known. Considering all 18 possible ratios, we get an average value of 
$(t_3/t_2)\sqrt{B_2/B_3}= 5.876$, leading to $B_2 / B_3=0.65$. 

By now the $E3$ transition matrix elements have been completely determined. The overall constant
\begin{equation}
F' = t_3 \sqrt{B/B_3}
\end{equation}
appearing in this case is connected to the overall constant $F$ through 
\begin{equation}
{F'\over F} = {t_3\over t_2} \sqrt{B_2\over B_3} = R \sqrt{B_2\over B_3},
\end{equation}
leading to $F'= 747.47$ and to the $E3$ matrix elements given in Table III. 

\subsubsection{$E1$ transitions}

In the case of $E1$ matrix elements, the quantity 
\begin{equation}
F''={1\over 2} t_1 {B \over \sqrt{B_2 B_3}} {e^{-{1\over  b^2}} \sin 2\phi_0 \over \sqrt{1- e^{-2b^2 \phi_0^2}}}
\end{equation}
can be treated as an overall constant, determined in the case of $^{224}$Ra by rms fitting to the two known transitions 
to be $F''=2.676\ 10^{-3}$ and providing the predictions given in Table III.

\section{Conclusions}

The analytic quadrupole octupole axially symmetric model with an infinite well potential (AQOA-IW) 
had successfully predicted the border between the regions of octupole deformation and octupole vibrations
in the light actinides, identifying $^{226}$Ra and $^{226}$Th as border nuclei \cite{AQOA}, with heavier isotopes 
corresponding to octupole vibrations and lighter isotopes exhibiting octupole deformation. The AQOA-IW model involved 
only one free parameter, $\phi_0$, expressing the relative presence of quadrupole vs. octupole deformation, while a parameter-free 
version has also been developed later \cite{Lenis}.

In the present work, the infinite well potential is substituted by a Davidson potential, resulting in the AQOA-D model,
which is able to deviate from the border line into the region of octupole deformation. This is achieved through the extra 
parameter $\beta_0$, the position of the minimum of the Davidson potential, which is increasing with increasing $R_{4/2}$ ratios,
as it is known from its use in the description of quadrupole deformed nuclei \cite{ESD}. 

Within the AQOA-D model, analytic expressions for energy spectra and B(E1), B(E2), B(E3) transition rates are derived. Then the following path 
is taken.

a) The spectra of $^{222-226}$Ra and $^{224,226}$Th [normalized to $E(2_1^+)$] are well reproduced in terms of the above mentioned two parameters $\phi_0$ and $\beta_0$.

b) The parameter $b$, related to the harmonic oscillator potential used in the $\phi$ degree of freedom, 
can be determined from the ratio of any $E2$ matrix element between negative parity states over any $E2$ matrix element between positive parity states,
fixing the determination of all $E2$ transitions up to an overall scale factor.

c) The ratio of mass parameters $B_2/B_3$ can be determined from the ratio of any $E3$ matrix element over any $E2$ matrix element, while 
the ratio of transition coefficients $t_2/t_3$ is fixed by the nuclear radius. As a result, the determination of all $E2$ and $E3$ transitions
is fixed, without any additional overall scale factor.

d) $E1$ transitions are also fixed, up to another scale factor. 

The recently measured $B(EL)$ transition rates of $^{224}$Ra \cite{224Ra}, presenting stable octupole deformation, 
provide a successful test for the model. It is clear that for other nuclei, the minimum set of data needed includes

a) A few energy levels of both positive and negative parity, from which the parameters $\phi_0$ and $\beta_0$ can be determined.

b) At least one $E2$ transition between positive parity states and one $E2$ transition between negative parity states,
from which the parameter $b$ can be determined.

c) At least one $E3$ transition, from which, in combination with the $E2$ transitions of b), the parameter ratio $B_2/B_3$ 
can be determined.  

From these pieces of data

a) The spectrum (leaving out the $\gamma$ bands) is determined up to an overall scale factor.

b) All relevant $E2$ and $E3$ transitions are determined up to an overall scale factor.

c) All relevant $E1$ transitions are determined up to another overall scale factor.

It is of interest to apply the present model in the actinides close to $^{240}$Pu, in which a second order 
shape phase transition from octupole-nondeformed to octupole-deformed shapes has been recently found \cite{Jolos86},
while octupole bands have been described \cite{Jolos88} using supersymmetric quantum mechanics.
The light rare earths, in which octupole bands have been considered recently both by the Bizzeti and Bizzeti-Sona approach 
\cite{Bizzeti81} and within density functional theory \cite{Guzman}, are also of special interest. 
A successful application of the AQOA-IW model to $^{148}$Nd has already been given in Ref. \cite{Sugawara}.  

\section*{Acknowledgements}

Financial support from the Bulgarian National Science Fund under contract No. DFNI-E02/6 and by
the Scientific Research Projects Coordination Unit of Istanbul University under Project No 50822
is gratefully acknowledged. 

\section*{Appendix A. $\phi$ integrals}

Since we confine ourselves to states with $n_\phi=0$, this quantum number 
is omitted in the notation of the wave functions, which then carry only the subscript 
$i$ ($f$) for the initial (final) state. 

For symmetric states one has 
\begin{equation}
X_S(\phi) =N_S {\chi(\tilde \phi^+)+\chi(\tilde \phi^-) \over \sqrt{2}}, 
\end{equation}
while for antisymmetric states one has 
\begin{equation}
X_A(\phi) =N_A {\chi(\tilde \phi^+)-\chi(\tilde \phi^-) \over \sqrt{2}}, 
\end{equation}
where $N_S$ and $N_A$ are normalization factors and, according to Eq. (\ref{eq:e22}),  
\begin{multline}
\chi(\tilde \phi^+)= \sqrt{b\over \sqrt{\pi}} e^{-{b^2\over 2}(\phi-\phi_0)^2}, \\ 
\chi(\tilde \phi^-)= \sqrt{b\over \sqrt{\pi}} e^{-{b^2\over 2}(\phi+\phi_0)^2}. 
\end{multline}

\subsection*{A0. Normalization}

For symmetric states one has 
\begin{multline}
{1\over N_S^2} = \int_{-\infty}^{\infty} X_S  X_S d\phi \\
 = {1\over 2} \int_{-\infty}^{\infty} (\chi(\phi^+))^2  d\phi
+ {1\over 2} \int_{-\infty}^{\infty} (\chi(\phi^-))^2  d\phi \\ + 
 \int_{-\infty}^{\infty} \chi(\phi^+) \chi(\phi^-)  d\phi \\ =
{1\over 2} {b\over \sqrt{\pi}} e^{-b^2 \phi_0^2} \int_{-\infty}^{\infty} e^{-b^2 \phi^2 + 2b^2 \phi_0 \phi}  d\phi \\ 
+ {1\over 2} {b\over \sqrt{\pi}} e^{-b^2 \phi_0^2} \int_{-\infty}^{\infty} e^{-b^2 \phi^2 - 2b^2 \phi_0 \phi}  d\phi \\
+ {b\over \sqrt{\pi}} e^{-b^2 \phi_0^2} \int_{-\infty}^{\infty} e^{-b^2 \phi^2} d\phi. 
\end{multline}
Using Eq. (\ref{elem}) of Appendix A4 we see that the integrals appearing here are of the form 
\begin{equation}
\int_{-\infty}^{\infty} e^{-b^2 \phi^2 \pm  2b^2 \phi_0 \phi}  d\phi = {\sqrt{\pi}\over b} e^{b^2 \phi_0^2},  
\end{equation}
leading to
\begin{equation}
{1\over N_S^2}= 1+ e^{-b^2 \phi_0^2}.
\end{equation}

For antisymmetric states one has 
\begin{multline}
{1\over N_A^2} = \int_{-\infty}^{\infty} X_A X_A d\phi \\
 = {1\over 2} \int_{-\infty}^{\infty} (\chi(\phi^+))^2  d\phi + {1\over 2} \int_{-\infty}^{\infty} (\chi(\phi^-))^2  d\phi \\- 
 \int_{-\infty}^{\infty} \chi(\phi^+) \chi(\phi^-)  d\phi \\ =
{1\over 2} {b\over \sqrt{\pi}} e^{-b^2 \phi_0^2} \int_{-\infty}^{\infty} e^{-b^2 \phi^2 + 2b^2 \phi_0 \phi}  d\phi \\
+ {1\over 2} {b\over \sqrt{\pi}} e^{-b^2 \phi_0^2} \int_{-\infty}^{\infty} e^{-b^2 \phi^2 - 2b^2 \phi_0 \phi}  d\phi \\
- {b\over \sqrt{\pi}} e^{-b^2 \phi_0^2} \int_{-\infty}^{\infty} e^{-b^2 \phi^2}  d\phi, 
\end{multline}
leading in the same way as above to 
\begin{equation}
{1\over N_A^2}= 1- e^{-b^2 \phi_0^2}. 
\end{equation}

\subsection*{A1. $B(E2)$s}

The transition operator for 
$B(E2)$s contains $\tilde \beta_2 = \tilde \beta \cos\phi$.  

For $B(E2)$s between symmetric states one has 
\begin{multline}
I_{\phi,S\to S}^{(E2)} = \int_{-\infty}^{\infty} X_S \cos\phi X_S d\phi \\
 = {N_S^2\over 2} \int_{-\infty}^{\infty} (\chi(\phi^+))^2 \cos\phi d\phi \\
+ {N_S^2\over 2} \int_{-\infty}^{\infty} (\chi(\phi^-))^2 \cos\phi d\phi \\ + 
N_S^2 \int_{-\infty}^{\infty} \chi(\phi^+) \chi(\phi^-) \cos\phi d\phi \\ =
{N_S^2\over 2} {b\over \sqrt{\pi}} e^{-b^2 \phi_0^2} \int_{-\infty}^{\infty} e^{-b^2 \phi^2 + 2b^2 \phi_0 \phi} \cos\phi d\phi \\ 
+ {N_S^2\over 2} {b\over \sqrt{\pi}} e^{-b^2 \phi_0^2} \int_{-\infty}^{\infty} e^{-b^2 \phi^2 - 2b^2 \phi_0 \phi} \cos\phi d\phi \\
+ N_S^2{b\over \sqrt{\pi}} e^{-b^2 \phi_0^2} \int_{-\infty}^{\infty} e^{-b^2 \phi^2} \cos\phi d\phi. 
\end{multline}
Using Eq. (\ref{INT1}) of Appendix A4 we see that the integrals appearing here are of the form 
\begin{equation}
\int_{-\infty}^{\infty} e^{-b^2 \phi^2 \pm 2b^2 \phi_0 \phi} \cos\phi d\phi = {\sqrt{\pi}\over b} e^{b^2 \phi_0^2 -{1\over 4 b^2}} \cos\phi_0,  
\end{equation}
leading to
\begin{equation}
I_{\phi,S\to S}^{(E2)}= e^{-{1\over 4b^2}} { \cos\phi_0+ e^{-b^2 \phi_0^2} \over 1+ e^{-b^2 \phi_0^2}}. 
\end{equation}

For $B(E2)$s between antisymmetric states one has 
\begin{multline}
I_{\phi,A\to A}^{(E2)} = \int_{-\infty}^{\infty} X_A \cos\phi X_A d\phi \\
 = {N_A^2\over 2} \int_{-\infty}^{\infty} (\chi(\phi^+))^2 \cos\phi d\phi \\
 + {N_A^2\over 2} \int_{-\infty}^{\infty} (\chi(\phi^-))^2 \cos\phi d\phi \\- 
N_A^2 \int_{-\infty}^{\infty} \chi(\phi^+) \chi(\phi^-) \cos\phi d\phi \\ =
{N_A^2\over 2} {b\over \sqrt{\pi}} e^{-b^2 \phi_0^2} \int_{-\infty}^{\infty} e^{-b^2 \phi^2 + 2b^2 \phi_0 \phi} \cos\phi d\phi \\
+ {N_A^2\over 2} {b\over \sqrt{\pi}} e^{-b^2 \phi_0^2} \int_{-\infty}^{\infty} e^{-b^2 \phi^2 - 2b^2 \phi_0 \phi} \cos\phi d\phi \\
- N_A^2{b\over \sqrt{\pi}} e^{-b^2 \phi_0^2} \int_{-\infty}^{\infty} e^{-b^2 \phi^2} \cos\phi d\phi, 
\end{multline}
leading in the same way as above to 
\begin{equation}
I_{\phi,A\to A}^{(E2)}= e^{-{1\over 4b^2}} { \cos\phi_0- e^{-b^2 \phi_0^2} \over 1- e^{-b^2 \phi_0^2}}. 
\end{equation}

For $B(E2)$s between symmetric and antisymmetric states one has 
\begin{multline}
I_{\phi,S\to A}^{(E2)} = \int_{-\infty}^{\infty} X_S \cos\phi X_A d\phi \\
= {N_S N_A \over 2} \int_{-\infty}^{\infty} (\chi(\phi^+))^2 \cos\phi d\phi \\
- {N_S N_A \over 2} \int_{-\infty}^{\infty} (\chi(\phi^-))^2 \cos\phi d\phi \\ =
{N_S N_A\over 2} {b\over \sqrt{\pi}} e^{-b^2 \phi_0^2} \int_{-\infty}^{\infty} e^{-b^2 \phi^2 + 2b^2 \phi_0 \phi} \cos\phi d\phi \\ 
- {N_S N_A\over 2} {b\over \sqrt{\pi}} e^{-b^2 \phi_0^2} \int_{-\infty}^{\infty} e^{-b^2 \phi^2 - 2b^2 \phi_0 \phi} \cos\phi d\phi, 
\end{multline}
leading to 
\begin{equation}
I_{\phi,S\to A}^{(E2)}= 0. 
\end{equation}
In the same way one also finds
\begin{equation}
I_{\phi,A\to S}^{(E2)}= 0. 
\end{equation}

\subsection*{A2. $B(E3)$s}

The transition operator for 
$B(E3)$s contains $\tilde \beta_3 = \tilde \beta \sin\phi$.  

For $B(E3)$s between symmetric states one has 
\begin{multline}
I_{\phi,S\to S}^{(E3)} = \int_{-\infty}^{\infty} X_S \sin\phi X_S d\phi \\
= {N_S^2\over 2} \int_{-\infty}^{\infty} (\chi(\phi^+))^2 \sin\phi d\phi \\
+ {N_S^2\over 2} \int_{-\infty}^{\infty} (\chi(\phi^-))^2 \sin\phi d\phi \\ + 
N_S^2 \int_{-\infty}^{\infty} \chi(\phi^+) \chi(\phi^-) \sin\phi d\phi \\ =
{N_S^2\over 2} {b\over \sqrt{\pi}} e^{-b^2 \phi_0^2} \int_{-\infty}^{\infty} e^{-b^2 \phi^2 + 2b^2 \phi_0 \phi} \sin\phi d\phi \\ 
+ {N_S^2\over 2} {b\over \sqrt{\pi}} e^{-b^2 \phi_0^2} \int_{-\infty}^{\infty} e^{-b^2 \phi^2 - 2b^2 \phi_0 \phi} \sin\phi d\phi \\
+ N_S^2 {b\over \sqrt{\pi}} e^{-b^2 \phi_0^2} \int_{-\infty}^{\infty} e^{-b^2 \phi^2} \sin\phi d\phi. 
\end{multline}
Using Eq. (\ref{INT2}) of Appendix A4 we see that the integrals appearing here are of the form 
\begin{equation}
\int_{-\infty}^{\infty} e^{-b^2 \phi^2 \pm  2b^2 \phi_0 \phi} \sin\phi d\phi = \pm {\sqrt{\pi}\over b} e^{b^2 \phi_0^2 -{1\over 4 b^2}} \sin\phi_0,  
\end{equation}
leading to
\begin{equation}
I_{\phi,S\to S}^{(E3)}= 0. 
\end{equation}

For $B(E3)$s between antisymmetric states one has 
\begin{multline}
I_{\phi,A\to A}^{(E3)} = \int_{-\infty}^{\infty} X_A \sin\phi X_A d\phi \\
= {N_S^2\over 2} \int_{-\infty}^{\infty} (\chi(\phi^+))^2 \sin\phi d\phi \\
+ {N_S^2\over 2} \int_{-\infty}^{\infty} (\chi(\phi^-))^2 \sin\phi d\phi \\ - 
N_S^2 \int_{-\infty}^{\infty} \chi(\phi^+) \chi(\phi^-) \sin\phi d\phi \\ =
{N_S^2\over 2} {b\over \sqrt{\pi}} e^{-b^2 \phi_0^2} \int_{-\infty}^{\infty} e^{-b^2 \phi^2 + 2b^2 \phi_0 \phi} \sin\phi d\phi \\ 
+ {N_S^2\over 2} {b\over \sqrt{\pi}} e^{-b^2 \phi_0^2} \int_{-\infty}^{\infty} e^{-b^2 \phi^2 - 2b^2 \phi_0 \phi} \sin\phi d\phi \\
- N_S^2 {b\over \sqrt{\pi}} e^{-b^2 \phi_0^2} \int_{-\infty}^{\infty} e^{-b^2 \phi^2} \sin\phi d\phi, 
\end{multline}
leading in the same way as above to 
\begin{equation}
I_{\phi,A\to A}^{(E3)}= 0. 
\end{equation}

For $B(E3)$s between symmetric and antisymmetric states one has 
\begin{multline}
I_{\phi,S\to A}^{(E3)} = \int_{-\infty}^{\infty} X_S \sin\phi X_A d\phi \\
= {N_S N_A\over 2} \int_{-\infty}^{\infty} (\chi(\phi^+))^2 \sin\phi d\phi \\
- {N_S N_A\over 2} \int_{-\infty}^{\infty} (\chi(\phi^-))^2 \sin\phi d\phi \\ =
{N_S N_A\over 2} {b\over \sqrt{\pi}} e^{-b^2 \phi_0^2} \int_{-\infty}^{\infty} e^{-b^2 \phi^2 + 2b^2 \phi_0 \phi} \sin\phi d\phi \\
- {N_S N_A\over 2} {b\over \sqrt{\pi}} e^{-b^2 \phi_0^2} \int_{-\infty}^{\infty} e^{-b^2 \phi^2 - 2b^2 \phi_0 \phi} \sin\phi d\phi, 
\end{multline}
leading to 
\begin{equation}
I_{\phi,S\to A}^{(E3)}= {e^{-{1\over 4 b^2}} \sin\phi_0 \over \sqrt{1- e^{-2b^2 \phi_0^2}}}. 
\end{equation}
In the same way one finds
\begin{equation}
I_{\phi,A\to S}^{(E3)}= I_{\phi,S\to A}^{(E3)}.  
\end{equation} 

\subsection*{A3. $B(E1)$s}

The transition operator for 
$B(E1)$s contains $\tilde \beta_2 \tilde \beta_3 = \tilde \beta \cos\phi \tilde \beta \sin \phi = \tilde\beta^2 {\sin 2\phi \over 2}$.  

For $B(E1)$s between symmetric states one has 
\begin{multline}
I_{\phi,S\to S}^{(E1)} = {1\over 2} \int_{-\infty}^{\infty} X_S \sin2\phi X_S d\phi\\
 = {N_S^2\over 4} \int_{-\infty}^{\infty} (\chi(\phi^+))^2 \sin2\phi d\phi \\
 + {N_S^2\over 4} \int_{-\infty}^{\infty} (\chi(\phi^-))^2 \sin2\phi d\phi \\
+ {N_S^2\over 2} \int_{-\infty}^{\infty} \chi(\phi^+) \chi(\phi^-) \sin2\phi d\phi \\ =
{N_S^2\over 4} {b\over \sqrt{\pi}} e^{-b^2 \phi_0^2} \int_{-\infty}^{\infty} e^{-b^2 \phi^2 + 2b^2 \phi_0 \phi} \sin2\phi d\phi \\
+ {N_S^2\over 4} {b\over \sqrt{\pi}} e^{-b^2 \phi_0^2} \int_{-\infty}^{\infty} e^{-b^2 \phi^2 - 2b^2 \phi_0 \phi} \sin2\phi d\phi \\
+ {N_S^2\over 2} {b\over \sqrt{\pi}} e^{-b^2 \phi_0^2} \int_{-\infty}^{\infty} e^{-b^2 \phi^2} \sin2\phi d\phi. 
\end{multline}
Using Eq. (\ref{INT2}) of Appendix A4 we see that the integrals appearing here are of the form 
\begin{equation}
\int_{-\infty}^{\infty} e^{-b^2 \phi^2 \pm  2b^2 \phi_0 \phi} \sin2\phi d\phi = \pm {\sqrt{\pi}\over b} e^{b^2 \phi_0^2 -{1\over b^2}} \sin2\phi_0,  
\end{equation}
leading to
\begin{equation}
I_{\phi,S\to S}^{(E1)}= 0. 
\end{equation}

For $B(E1)$s between antisymmetric states one has 
\begin{multline}
I_{\phi,A\to A}^{(E1)} = {1\over 2}\int_{-\infty}^{\infty} X_A \sin2\phi X_A d\phi \\
= {N_A^2\over 4} \int_{-\infty}^{\infty} (\chi(\phi^+))^2 \sin2\phi d\phi \\
+ {N_A^2\over 4} \int_{-\infty}^{\infty} (\chi(\phi^-))^2 \sin2\phi d\phi \\
- {N_A^2\over 2}\int_{-\infty}^{\infty} \chi(\phi^+) \chi(\phi^-) \sin2\phi d\phi \\ =
{N_A^2\over 4} {b\over \sqrt{\pi}} e^{-b^2 \phi_0^2} \int_{-\infty}^{\infty} e^{-b^2 \phi^2 + 2b^2 \phi_0 \phi} \sin2\phi d\phi \\
+ {N_A^2\over 4} {b\over \sqrt{\pi}} e^{-b^2 \phi_0^2} \int_{-\infty}^{\infty} e^{-b^2 \phi^2 - 2b^2 \phi_0 \phi} \sin2\phi d\phi \\
- {N_A^2\over 2} {b\over \sqrt{\pi}} e^{-b^2 \phi_0^2} \int_{-\infty}^{\infty} e^{-b^2 \phi^2} \sin2\phi d\phi, 
\end{multline}
leading in the same way as above to 
\begin{equation}
I_{\phi,A\to A}^{(E1)}= 0. 
\end{equation}

For $B(E1)$s between symmetric and antisymmetric states one has 
\begin{multline}
I_{\phi,S\to A}^{(E1)} ={1\over 2} \int_{-\infty}^{\infty} X_S \sin2\phi X_A d\phi \\
= {N_S N_A \over 4} \int_{-\infty}^{\infty} (\chi(\phi^+))^2 \sin2\phi d\phi \\
- {N_S N_A\over 4} \int_{-\infty}^{\infty} (\chi(\phi^-))^2 \sin2\phi d\phi \\ =
{N_S N_A\over 4} {b\over \sqrt{\pi}} e^{-b^2 \phi_0^2} \int_{-\infty}^{\infty} e^{-b^2 \phi^2 + 2b^2 \phi_0 \phi} \sin2\phi d\phi \\ 
- {N_S N_A\over 4} {b\over \sqrt{\pi}} e^{-b^2 \phi_0^2} \int_{-\infty}^{\infty} e^{-b^2 \phi^2 - 2b^2 \phi_0 \phi} \sin2\phi d\phi, 
\end{multline}
leading to 
\begin{equation}
I_{\phi,S\to A}^{(E1)}=  {e^{-{1\over b^2}} \sin2\phi_0 \over 2 \sqrt{1- e^{-2b^2 \phi_0^2}}}. 
\end{equation}
In the same way one finds
\begin{equation}
I_{\phi,A\to S}^{(E1)}= I_{\phi,S\to A}^{(E1)}.  
\end{equation} 

\subsection*{A4. Useful integrals}

We know that (Eq. 3.897.2 of Ref. \cite{Grad})
\begin{multline}
\int_0^\infty e^{-\beta x^2 -\gamma x} \cos  bx dx = \\
{1\over 4} \sqrt{\pi \over b} \left\{ e^{(\gamma -ib)^2/(4\beta)} 
\left[1-\Phi\left(\gamma-ib\over 2\sqrt{\beta}  \right)  \right] \right. \\ 
\left.  + e^{(\gamma +ib)^2/(4\beta)} 
\left[1-\Phi\left(\gamma+ib\over 2\sqrt{\beta}  \right)  \right]  
\right\},  \label{Int1}
\end{multline}
where $Re\beta>0$, $\quad b>0$, and $\Phi(x)$ is the error function, having the property 
\begin{equation}\label{minus}
\Phi(-x)=-\Phi(x). 
\end{equation}
Changing the variable into $y=-x$, Eq. (\ref{Int1}) takes the form
\begin{multline}
-\int_0^{-\infty} e^{-\beta y^2 +\gamma y} \cos  by dy = \\
{1\over 4} \sqrt{\pi \over b} \left\{ e^{(\gamma -ib)^2/(4\beta)} 
\left[1-\Phi\left(\gamma-ib\over 2\sqrt{\beta}  \right)  \right] \right. \\ \left. +
e^{(\gamma +ib)^2/(4\beta)} 
\left[1-\Phi\left(\gamma+ib\over 2\sqrt{\beta}  \right)  \right]  
\right\}, 
\end{multline}
Changing the symbol $y$ into $x$ and letting $\gamma \to -\gamma$, one then gets 
\begin{multline}
\int_{-\infty}^0 e^{-\beta x^2 -\gamma x} \cos bx dx =  \\
{1\over 4} \sqrt{\pi \over b} \left\{ e^{(-\gamma -ib)^2/(4\beta)} 
\left[1-\Phi\left(-\gamma-ib\over 2\sqrt{\beta}  \right)  \right] \right. \\ \left. +
e^{(-\gamma +ib)^2/(4\beta)} 
\left[1-\Phi\left(-\gamma+ib\over 2\sqrt{\beta}  \right)  \right]  
\right\}.  \label{Int1m}
\end{multline}
Taking into account the property (\ref{minus}), Eq. (\ref{Int1m}) takes the form
\begin{multline}
\int_{-\infty}^0 e^{-\beta x^2 -\gamma x} \cos bx dx =  \\
{1\over 4} \sqrt{\pi \over b} \left\{ e^{(\gamma +ib)^2/(4\beta)} 
\left[1+\Phi\left(\gamma+ib\over 2\sqrt{\beta}  \right)  \right] \right. \\ \left. +
e^{(\gamma -ib)^2/(4\beta)} 
\left[1+\Phi\left(\gamma-ib\over 2\sqrt{\beta}  \right)  \right]  
\right\},  \label{Int1b}
\end{multline}
Adding Eqs. (\ref{Int1}) and (\ref{Int1b}), we get 
\begin{multline}\label{INT1}
\int_{-\infty}^\infty e^{-\beta x^2 -\gamma x} \cos  bx dx = {1\over 2} \sqrt{\pi \over \beta}
\left( e^{(\gamma+ib)^2 \over 4\beta } + e^{(\gamma-ib)^2 \over 4\beta }  \right) \\
= \sqrt{\pi\over \beta} e^{\gamma^2 -b^2 \over 4\beta} \cos{\gamma b \over 2\beta}.
\end{multline}

In a similar way, starting from the integral (Eq. 3.897.1 of Ref. \cite{Grad})
\begin{multline}
\int_0^\infty e^{-\beta x^2 -\gamma x} \sin  bx dx = \\
-{i\over 4} \sqrt{\pi \over b} \left\{ e^{(\gamma -ib)^2/(4\beta)} 
\left[1-\Phi\left(\gamma-ib\over 2\sqrt{\beta}  \right)  \right] \right. \\ \left. -
e^{(\gamma +ib)^2/(4\beta)} 
\left[1-\Phi\left(\gamma+ib\over 2\sqrt{\beta}  \right)  \right]  
\right\},  \label{Int2}
\end{multline}
where $Re\beta>0$, $\quad b>0$, we get
\begin{multline}\label{INT2} 
\int_{-\infty}^\infty e^{-\beta x^2 -\gamma x} \sin  bx dx = {i\over 2} \sqrt{\pi \over \beta}
\left( e^{(\gamma+ib)^2 \over 4\beta } - e^{(\gamma-ib)^2 \over 4\beta }  \right) \\
=- \sqrt{\pi\over \beta} e^{\gamma^2 -b^2 \over 4\beta} \sin{\gamma b \over 2\beta}.
\end{multline}

For normalization purposes the integral
\begin{equation}\label{elem}
\int_{-\infty}^\infty e^{-(ax^2+bx+c)} dx=\sqrt{\pi\over a} e^{(b^2-4ac)/(4a)} 
\end{equation}
suffices. 

\section*{Appendix B. $\theta$ integrals}

Integrals over $\theta$ involve three Wigner functions and can be calculated using 
Eq. (4.6.2) of Ref. \cite{Edmonds}
\begin{multline}
\int d^3\theta  {\cal D}^{(j_1)}_{k_1 m_1}(\theta)  {\cal D}^{(j_2)}_{k_2 m_2}(\theta) {\cal D}^{(j_3)}_{k_3 m_3}(\theta) \\=
8\pi^2 \left(\begin{array}{ccc}  j_1 & j_2 & j_3 \\ k_1 & k_2 & k_3 \end{array} \right) 
\left(\begin{array}{ccc}  j_1 & j_2 & j_3 \\ m_1 & m_2 & m_3 \end{array} \right) . 
\end{multline}
Using the relation between 3-j symbols and Clebsch Gordan coefficients (3.7.3) of \cite{Edmonds}
\begin{equation}
\left(\begin{array}{ccc}  j_1 & j_2 & j_3 \\ m_1 & m_2 & m_3 \end{array}\right) = {(-1)^{j_1-j_2-m_3} \over \sqrt{2j_3+1}} (j_1 j_2 j_3 | m_1 m_2 -m_3),  
\end{equation} 
and the relation for conjugate Wigner functions (4.2.7) of \cite{Edmonds}
\begin{equation}
{\cal D}^{(j)*}_{km}(\theta) = (-1)^{k-m} {\cal D}^{(j)}_{-k-m}(\theta),
\end{equation}
one obtains 
\begin{multline}
\int d^3\theta  {\cal D}^{(j_1)}_{k_1 m_1}(\theta)  {\cal D}^{(j_2)}_{k_2 m_2}(\theta) {\cal D}^{(j_3)*}_{-k_3-m_3}(\theta) (-1)^{2k_3} \\ =
{8\pi^2 \over 2j_3+1} ( j_1  j_2  j_3 | k_1  k_2  -k_3) (j_1  j_2  j_3 | m_1  m_2  -m_3),  
\end{multline}
which by replacing $m_3$ ($k_3$) by $-m_3$ ($-k_3$) can be rewritten as
\begin{multline}\label{Edm}
\int d^3\theta  {\cal D}^{(j_1)}_{k_1 m_1}(\theta)  {\cal D}^{(j_2)}_{k_2 m_2}(\theta) {\cal D}^{(j_3)*}_{k_3 m_3}(\theta) (-1)^{-2k_3} \\ =
{8\pi^2 \over 2j_3+1} ( j_1  j_2  j_3 | k_1  k_2  k_3) ( j_1  j_2  j_3 | m_1  m_2  m_3).   
\end{multline}

\subsection*{B1. $B(E2)$s} 

In this case the integral reads 
\begin{multline}
I_\theta^{(E2)} = \int d^3\theta \sqrt{2L_i+1\over 32\pi^2} [1\pm (-1)^{L_i}]{\cal D}^{L_i}_{0,M_i}(\theta) {\cal D}^{(2)}_{0,\mu}(\theta) \\
\sqrt{2L_f+1\over 32\pi^2} [1\pm (-1)^{L_f}]{\cal D}^{L_f*}_{0,M_f}(\theta). 
\end{multline}
Using Eq. (\ref{Edm}) this gives 
\begin{multline}\label{result}
I_\theta^{(E2)} = {(1 \pm (-1)^{L_i})(1 \pm (-1)^{L_f}) \over 4} 
\sqrt{ 2L_i+1 \over 2L_f+1} \\ (L_i 2 L_f | 0 0 0) 
(L_i 2 L_f | M_i \mu M_f) . 
\end{multline}

From the $\phi$ integrals we know that non-vanishing results are obtained only in the $S\to S$ and $A\to A$ cases. 

In the $S\to S$ case the two factors in the rhs of Eq. (\ref{result}) have the positive signs in the place of the double signs, 
thus allowing only even values of $L_i$ and $L_f$, resulting in a factor of 4 in the numerator. 

In the $A\to A$ case the two factors in the rhs of Eq. (\ref{result}) have the negative signs in the place of the double signs, 
thus allowing only odd values of $L_i$ and $L_f$, resulting again in a factor of 4 in the numerator. 

As a consequence, in all cases the final reasult reads 
\begin{equation}
I_\theta^{(E2)} =  \sqrt{ 2L_i+1 \over 2L_f+1} (L_i 2 L_f | 0 0 0) (L_i 2 L_f | M_i \mu M_f) . 
\end{equation}
 
\subsection*{B2. $B(E3)$s}

The calculation parallels the one of the previous subsection, the only difference being that 
the middle term, coming from the transition operator, is ${\cal D}^{(3)}_{0,\mu}$. The result reads
\begin{multline}\label{res3}
I_\theta^{(E3)} ={(1 \pm (-1)^{L_i})(1 \pm (-1)^{L_f}) \over 4} 
\sqrt{ 2L_i+1 \over 2L_f+1} \\ (L_i 3 L_f | 0 0 0) 
(L_i 3 L_f | M_i \mu M_f) . 
\end{multline}

From the $\phi$ integrals we know that non-vanishing results are obtained only in the $S\to A$ and $A\to S$ cases. 

In the $S\to A$ case, $L_i$ is even and $L_f$ is odd.  The first factor in the rhs of Eq. (\ref{res3}) has the positive sign in the place of the double sign in front of the $(-1)^{L_i}$ term and the negative sign in the place of the double sign in front of the $(-1)^{L_f}$ term,
resulting in a factor of 4 in the numerator. The same factor of 4 is obtained also in the $A\to S$ case. Therefore in all cases the final result reads 
 \begin{equation}
I_\theta^{(E3)} =  \sqrt{ 2L_i+1 \over 2L_f+1} (L_i 3 L_f | 0 0 0) (L_i 3 L_f | M_i \mu M_f) . 
\end{equation}

\subsection*{B3. $B(E1)$s}  

The calculation parallels the one of the previous subsection, the only difference being that 
the middle term, coming from the transition operator, is ${\cal D}^{(1)}_{0,\mu}$. The result reads
\begin{multline}\label{resul3}
I_\theta^{(E1)} = {(1 \pm (-1)^{L_i})(1 \pm (-1)^{L_f}) \over 4} 
\sqrt{ 2L_i+1 \over 2L_f+1} \\ (L_i 1 L_f | 0 0 0) 
(L_i 1 L_f | M_i \mu M_f) . 
\end{multline}

From the $\phi$ integrals we know that non-vanishing results are obtained only in the $S\to A$ and $A\to S$ cases. 

In the $S\to A$ case, $L_i$ is even and $L_f$ is odd.  The first factor in the rhs of Eq. (\ref{resul3}) has the positive sign in the place of the double sign in front of the $(-1)^{L_i}$ term and the negative sign in the place of the double sign in front of the $(-1)^{L_f}$ term,
resulting in a factor of 4 in the numerator. The same factor of 4 is obtained also in the $A\to S$ case. Therefore in all cases the final result reads 
 \begin{equation}
I_\theta^{(E1)} =  \sqrt{ 2L_i+1 \over 2L_f+1} (L_i 1 L_f | 0 0 0) (L_i 1 L_f | M_i \mu M_f) . 
\end{equation}

\subsection*{B4. Normalization}

Normalization of the $\theta$ wave functions is guaranteed by the integral of Eq. (4.6.1) of Ref. \cite{Edmonds}
\begin{equation}\label{twoD}
\int d^3\theta  {\cal D}^{(j_1)*}_{k_1 m_1}(\theta)  {\cal D}^{(j_2)}_{k_2 m_2}(\theta)=\delta_{k_1 k_2} \delta_{m_1 m_2} \delta_{j_1 j_2} 
{8\pi^2 \over 2j_1+1}. 
\end{equation}
The normalization integral for any state reads
\begin{multline}
I_{norm}= \int d^3\theta \sqrt{2L+1\over 32\pi^2} [1\pm (-1)^{L}]{\cal D}^{L*}_{0,M}(\theta)  \\
\sqrt{2L+1\over 32\pi^2} [1\pm (-1)^{L}]{\cal D}^{L}_{0,M}(\theta). 
\end{multline} 
Using Eq. (\ref{twoD}) this gives 
\begin{equation}
I_{norm} = {1+(-1)^{2L} \pm 2 (-1)^{L} \over 4} =1, 
\end{equation}
since for symmetric states the positive sign appears in the place of the double sign 
and $L$ is even, while for antisymmetric states the negative sign appears in the place of the double sign 
and $L$ is odd. 

\section*{Appendix C. $\tilde \beta$ integrals}

\subsection*{C1. $B(E2)$s} 

The transition operator contains a $\tilde \beta$ factor, thus the integrals appearing in this case read
\begin{equation}\label{eq:e39}
I_{\tilde \beta}^{(E2)} = \sqrt{{2 (n_i)! \over \Gamma(n_i+a_i+1)} {2 (n_f)! \over \Gamma(n_f+a_f+1)}} I^{(E2)}(n_i,n_f),
\end{equation}
with
\begin{equation}
I^{(E2)}(n_i,n_f)= \int_0^\infty  \tilde \beta^{a_i+a_f+2}   e^{-\tilde \beta^2} L_{n_i}^{a_i}(\tilde \beta^2)
 L_{n_f}^{a_f}(\tilde \beta^2) d\tilde \beta . 
\end{equation}  
Using the substitution $\tilde \beta^{2}=x$ with $dx=2\tilde \beta d\tilde \beta$, the integral is written as 
\begin{equation}
I^{(E2)}(n_i,n_f) =\frac{1}{2}\int_0^\infty e^{-x} x^{\frac{a_i}{2}+\frac{a_f}{2}+\frac{1}{2}}
L_{n_i}^{a_i}(x) L_{n_f}^{a_f}(x) dx. 
\label{Ilx}
\end{equation}

Analytic expressions can be found for these integrals in the case in which one of the quantum numbers $n_i$, $n_f$ is zero.
(For the case in which both quantum numbers $n_i$ and $n_f$ are non-zero, see Eq. (B5) of Ref. \cite{Strecker}.)
We consider $n_f=0$, since both the ground state band and the octupole band are characterized by this value.  
Then one has $L_{n_{f}=0}^{a_{f}}(x)=1$ and the integral is simplified into 
\begin{equation}
I^{(E2)}(n_i,0) =\frac{1}{2}\int_{0}^{\infty} e^{-x}
x^{\frac{a_i}{2}+\frac{a_f}{2}+\frac{1}{2}} L_{n_i}^{a_i}(x)
 dx .
\label{Ilx0}
\end{equation}
Integrals of this form are known to have the following
analytic solution (\cite{Prudnikov}, p. 463, Eq. (5))
\begin{equation}
\int_{0}^{\infty}  e^{-cx} x^{(\alpha-1)}   L_{n}^{\lambda}(cx)  dx=  \frac{(1-\alpha
+\lambda)_{n}}{n!c^{\alpha}}\Gamma(\alpha) ,
\label{p463}
\end{equation}
where $(a)_{n}$ is the Pochhammer symbol
\begin{eqnarray}
(a)_{n}=\frac{\Gamma(a+n)}{\Gamma(a)}.
\label{Poch}
\end{eqnarray}
By replacing $c=1$,  $\lambda =a_{i}$, 
$\alpha-1=\frac{a_{i}}{2}+\frac{a_{f}}{2}+\frac{1}{2}$, i.e. $\alpha
=\frac{a_{i}}{2}+\frac{a_{f}}{2}+\frac{3}{2}$, and applying the definition
(\ref{Poch}) the result is
\begin{equation}
I^{(E2)}(n_i,0) =\frac{1}{2}\frac{\Gamma\left(n_{i}+\frac{a_{i}}{2}-\frac{a_{f}}{2}-\frac{1}{2}\right)}
{n_{i}!\Gamma\left(\frac{a_{i}}{2}-\frac{a_{f}}{2}-\frac{1}{2}\right)}
\Gamma\left(\frac{a_{i}}{2}+\frac{a_{f}}{2}+\frac{3}{2}\right).
\label{Ilx0a}
\end{equation}
In the simplest case of a transition between states with $n_i=0$ and $n_f=0$, which will
be eventually of major interest in the present work, one obviously has
\begin{equation}
I^{(E2)}(0,0) =\frac{1}{2}\Gamma\left(\frac{a_{i}}{2}+\frac{a_{f}}{2}+\frac{3}{2}\right).
\label{Ilx0ayr}
\end{equation}

Substituting these results in Eq. (\ref{eq:e39}), for the case with $n_i\geq 0$ and $n_f=0$ we find 
\begin{multline}
I_{\tilde \beta}^{(E2)}
= \\ \frac{\Gamma\left(\frac{a_{i}}{2}+\frac{a_{f}}{2}+\frac{3}{2}\right)
\Gamma\left(n_{i}+\frac{a_{i}}{2}-\frac{a_{f}}{2}-\frac{1}{2}\right)}
{\sqrt{n_{i}!\Gamma(n_{i}+a_{i}+1)\Gamma(a_{f}+1)}
\Gamma\left(\frac{a_{i}}{2}-\frac{a_{f}}{2}-\frac{1}{2}\right)},
\label{snf0}
\end{multline}
while in the simplest case of $n_i= 0$ and $n_f=0$ one has
\begin{eqnarray}
I_{\tilde \beta}^{(E2)}
=\frac{\Gamma\left(\frac{a_{i}}{2}+\frac{a_{f}}{2}+\frac{3}{2}\right)}
{\sqrt{\Gamma(a_{i}+1)\Gamma(a_{f}+1)}}.
\label{snf00}
\end{eqnarray}

In the case of $n_i=0$, $n_f\geq 0$, following the same steps one finds 
\begin{multline}
I^{(E2)}(0,n_f) =\\ \frac{1}{2}\frac{\Gamma\left(n_{f}+\frac{a_{f}}{2}-\frac{a_{i}}{2}-\frac{1}{2}\right)}
{n_{f}!\Gamma\left(\frac{a_{f}}{2}-\frac{a_{i}}{2}-\frac{1}{2}\right)}
\Gamma\left(\frac{a_{i}}{2}+\frac{a_{f}}{2}+\frac{3}{2}\right),
\end{multline}
\begin{multline}
I_{\tilde \beta}^{(E2)}
=\\ \frac{\Gamma\left(\frac{a_{i}}{2}+\frac{a_{f}}{2}+\frac{3}{2}\right)
\Gamma\left(n_{f}+\frac{a_{f}}{2}-\frac{a_{i}}{2}-\frac{1}{2}\right)}
{\sqrt{n_{f}!\Gamma(n_{f}+a_{f}+1)\Gamma(a_{i}+1)}
\Gamma\left(\frac{a_{f}}{2}-\frac{a_{i}}{2}-\frac{1}{2}\right)}.
\end{multline}

\subsection*{C2. $B(E3)$s}

The transition operator again contains a $\tilde \beta$ factor, thus the integrals appearing in this case
are exactly the same as in the previous subsection
\begin{equation}
I_{\tilde \beta}^{(E3)}= I_{\tilde \beta}^{(E2)}. 
\end{equation}

\subsection*{C3. $B(E1)$s}  

The transition operator contains a $\tilde \beta^2 $ factor, thus the integrals appearing in this case read
\begin{equation}\label{eq:e40}
I_{\tilde \beta}^{(E1)} = \sqrt{{2 (n_i)! \over \Gamma(n_i+a_i+1)} {2 (n_f)! \over \Gamma(n_f+a_f+1)}} I^{(E1)}(n_i,n_f),
\end{equation}
with
\begin{equation}
I^{(E1)}(n_i,n_f)= \int_0^\infty  \tilde \beta^{a_i+a_f+3}   e^{-\tilde \beta^2} L_{n_i}^{a_i}(\tilde \beta^2)
 L_{n_f}^{a_f}(\tilde \beta^2) d\tilde \beta . 
\end{equation}  
Using again the substitution $\tilde \beta^{2}=x$ with $dx=2\tilde \beta d\tilde \beta$, the integral is written as 
\begin{equation}
I^{(E1)}(n_i,n_f) =\frac{1}{2}\int_0^\infty e^{-x} x^{\frac{a_i}{2}+\frac{a_f}{2}+1}
L_{n_i}^{a_i}(x) L_{n_f}^{a_f}(x) dx. 
\end{equation}

For $n_f=0$ the integral is simplified into 
\begin{equation}
I^{(E1)}(n_i,0) =\frac{1}{2}\int_{0}^{\infty} e^{-x}
x^{\frac{a_i}{2}+\frac{a_f}{2}+1} L_{n_i}^{a_i}(x) dx .
\end{equation}
Using Eq. (\ref{p463}) with $c=1$,  $\lambda =a_{i}$, 
$\alpha-1=\frac{a_{i}}{2}+\frac{a_{f}}{2}+1$, i.e. $\alpha
=\frac{a_{i}}{2}+\frac{a_{f}}{2}+2$, and applying the definition
(\ref{Poch}) the result is
\begin{equation}
I^{(E1)}(n_i,0) =\frac{1}{2}\frac{\Gamma\left(n_{i}+\frac{a_{i}}{2}-\frac{a_{f}}{2}-1\right)}
{n_{i}!\Gamma\left(\frac{a_{i}}{2}-\frac{a_{f}}{2}-1\right)}
\Gamma\left(\frac{a_{i}}{2}+\frac{a_{f}}{2}+2\right).
\end{equation}
In the simplest case of $n_i=0$ and $n_f=0$ one has
\begin{equation}
I^{(E1)}(0,0) =\frac{1}{2}\Gamma\left(\frac{a_{i}}{2}+\frac{a_{f}}{2}+2\right).
\end{equation}

Substituting these results in Eq. (\ref{eq:e40}), for the case with $n_i\geq 0$ and $n_f=0$ we find 
\begin{multline}
I_{\tilde \beta}^{(E1)}
=\\ \frac{\Gamma\left(\frac{a_{i}}{2}+\frac{a_{f}}{2}+2\right)
\Gamma\left(n_{i}+\frac{a_{i}}{2}-\frac{a_{f}}{2}-1\right)}
{\sqrt{n_{i}!\Gamma(n_{i}+a_{i}+1)\Gamma(a_{f}+1)}
\Gamma\left(\frac{a_{i}}{2}-\frac{a_{f}}{2}-1\right)},
\end{multline}
while in the simplest case of $n_i= 0$ and $n_f=0$ one has
\begin{eqnarray}
I_{\tilde \beta}^{(E1)}
=\frac{\Gamma\left(\frac{a_{i}}{2}+\frac{a_{f}}{2}+2\right)}
{\sqrt{\Gamma(a_{i}+1)\Gamma(a_{f}+1)}}.
\end{eqnarray}

In the case of $n_i=0$, $n_f\geq 0$, following the same steps one finds 
\begin{equation}
I^{(E1)}(0,n_f) =\frac{1}{2}\frac{\Gamma\left(n_{f}+\frac{a_{f}}{2}-\frac{a_{i}}{2}-1\right)}
{n_{f}!\Gamma\left(\frac{a_{f}}{2}-\frac{a_{i}}{2}-1\right)}
\Gamma\left(\frac{a_{i}}{2}+\frac{a_{f}}{2}+2\right),
\end{equation}
\begin{multline}
I_{\tilde \beta}^{(E1)}
= \\ \frac{\Gamma\left(\frac{a_{i}}{2}+\frac{a_{f}}{2}+2\right)
\Gamma\left(n_{f}+\frac{a_{f}}{2}-\frac{a_{i}}{2}-1\right)}
{\sqrt{n_{f}!\Gamma(n_{f}+a_{f}+1)\Gamma(a_{i}+1)}
\Gamma\left(\frac{a_{f}}{2}-\frac{a_{i}}{2}-1\right)}.
\end{multline}

\subsection*{C4. Normalization}

The total wave functions are given in Eqs. (\ref{eq:e2}) and (\ref{eq:e33}). 
The integration over the Euler angles $\theta$ and the relevant normalization 
have been studied in Appendix B, 
while the rest of the integrations are performed over 
$\int\int \beta_2^3 d\beta_2 \beta_3^3 d\beta_3$, where the $\beta_2^3$, 
$\beta_3^3$ factors come from the volume element and cancel with the 
first factor of Eq. (\ref{eq:e33}). Using Eqs. (\ref{eq:e5}) and 
(\ref{eq:e7}), as well as the relevant Jacobian, one finds that the rest of the integrations
are over $\int\int d\beta_2 d\beta_3 = {B\over \sqrt{B_2 B_3}} \int\int \tilde \beta d\tilde \beta d\phi$. 
The integration over $\phi$ and the relevant normalization factors have been studied in Appendix A. 
We determine here the normalization factors related to the $\tilde \beta$ integration. 
We have
\begin{equation}\label{normbeta}
{1\over N^2_{\tilde \beta}} = {2 (n)! \over \Gamma(n+a+1)} {B\over \sqrt{B_2 B_3}} I(n,n)
\end{equation} 
with
\begin{equation}
I(n,n)= \int_0^\infty  \tilde \beta^{2a+1}   e^{-\tilde \beta^2} L_{n}^{a}(\tilde \beta^2)
 L_{n}^{a}(\tilde \beta^2) d\tilde \beta . 
\end{equation}  
Using the substitution $\tilde \beta^{2}=x$ with $dx=2\tilde \beta d\tilde \beta$, the integral is written as 
\begin{equation}
I(n,n) =\frac{1}{2}\int_0^\infty e^{-x} x^a
L_{n}^{a}(x) L_{n}^{a}(x) dx. 
\end{equation}
Considering the case with $n=0$, which is of interest here, 
this integral is of the form of Eq. (\ref{p463}) with $c=1$, $\alpha-1=a$, thus leading to 
\begin{equation}
I(0,0)= {\Gamma(a+1)\over 2}. 
\end{equation}
Then Eq. (\ref{normbeta}) for $n=0$ leads to
\begin{equation}
{1\over N^2_{\tilde \beta}}= {B\over \sqrt{B_2 B_3}}. 
\end{equation}
This result indicates that when calculating $\tilde\beta$-integrals in $B(EL)$s, the $N_{\tilde \beta}$ normalization factors 
cancel out with the ${B\over \sqrt{B_2 B_3}}$ factor appearing in the volume element and therefore do not affect the final results. 

\section*{Appendix D.}

\subsection*{D1. Kinetic energy and volume elements}

The expressions for the kinetic energy and the volume element depend on the dimensionality 
of the space considered. We distinguish three cases, with dimensionality five, four, and three respectively.  

1) In the usual Bohr Hamiltonian describing the quadrupole degree of freedom in the five-dimensional (5D) space 
of the collective variables $\beta$ and $\gamma$ and the three Euler angles ($\theta$, $\phi$, $\psi$), the kinetic
energy term reads  \cite{Bohr}
\begin{multline}
T_{(\beta,\gamma)vib} =\\ -{\hbar^2 \over 2B} \left[ {1\over \beta^4} {\partial \over
\partial \beta} \beta^4 {\partial \over \partial \beta} + {1\over
\beta^2 \sin 3\gamma} {\partial \over \partial \gamma} \sin 3
\gamma {\partial \over \partial \gamma}\right],
\label{TvibBohr}
\end{multline}
resulting from the Pauli--Podolsky quantization procedure \cite{Podolsky} in the full 5D space.
The volume element reads \cite{Bohr} 
\begin{equation}
d\tau= \beta^4 \vert \sin 3\gamma\vert \sin\theta \ d\beta \ d\gamma \ d\theta \ d\phi \ d\psi.
\end{equation}
If the $\beta$ variable is separated from the rest, either exactly, as in the E(5) critical point symmetry \cite{IacE5},
or through an adiabatic approximation, as in the X(5) approach \cite{IacX5}, 
the volume element in the $\beta$ part of the problem becomes \cite{IacE5,IacX5}
\begin{equation}
d\tau_{\beta}(\beta)= \beta^{4} d\beta .
\end{equation}

2) In the Davydov--Chaban approach \cite{Chaban} the $\gamma$ variable is removed from the
Hamiltonian from the very beginning of the problem, since $\gamma$ is treated as an
effective deformation parameter. Then the quantization procedure is applied in the
4D curvilinear space of $\beta$ and the three Euler angles. As a result the kinetic
energy term of the Hamiltonian is obtained in the form
\begin{equation}
T_{(\beta)vib} = -{\hbar^2 \over 2B} \left[ {1\over \beta^3} {\partial \over
\partial \beta} \beta^3 {\partial \over \partial \beta} \right].
\label{TvibDCh}
\end{equation}
Note that now the power of $\beta$ in (\ref{TvibDCh}) is 3 and not 4 as in the
$\beta$-part of (\ref{TvibBohr}), while the respective volume element is
\begin{equation}
d\tau_{(\beta)vib}(\beta)= \beta^{3}d\beta.
\end{equation}
If the wave function is sought in the form \cite{Chaban}
\begin{equation}
\psi (\beta)=\beta^{-3/2}\varphi (\beta),
\label{wfChaban}
\end{equation}
the kinetic energy term in the Schr\"odinger equation for the wave function $\varphi (\beta)$
appears in the form \cite{Chaban}
\begin{equation}
\widetilde{T}_{(\beta)vib} =-{\hbar^2\over 2 B_2} {\partial^2 \over \partial \beta_2^2}
+{ 3\hbar^2\over 8B_2 \beta_2^2} ,
\label{Tvibeff}
\end{equation}
where the second term is further moved into the effective potential part.

3) In the limit of strong $\gamma$ instability of the Wilets--Jean approach \cite{Wilets} the nucleus 
is considered as a droplet which can only execute axially symmetric vibrations. This system has only 
three degrees of freedom: $\beta$, $\theta$ and $\phi$. 
Then the kinetic energy term 
in the Hamiltonian becomes 
\begin{equation}
T = -{\hbar^2 \over 2B} {\partial^2\over \partial \beta^2}, 
\end{equation}
where wave functions of the form $\psi (\beta)=\beta^{-1}\varphi (\beta)$ are considered 
and the volume element reads
\begin{equation}
d\tau= \beta^2 \sin\theta \ d\beta \ d\theta \ d\phi. 
\end{equation}
This approach has been recently used in Ref. \cite{Shapirov}. 

The kinetic energy term of the Davydov--Chaban approach has been generalized from quadrupole to any multipolarity 
$\lambda$ by Williams and Davidson \cite{Williams}, the final result being 
\begin{equation}
T_{\lambda} = -{\hbar^2 \over 2B_\lambda} \left[ {1\over \beta_\lambda^3} {\partial \over
\partial \beta_\lambda} \beta_\lambda^3 {\partial \over \partial \beta_\lambda} \right].
\end{equation}
The basic assumption behind this derivation is the requirement of no vibration-rotation cross terms
\cite{Williams,Lipas}, which diagonalizes the inertial tensor and hence the rotated coordinate system 
is the principal inertial (body) system. 

In the case of simultaneous presence of quadrupole and octupole deformation, 
the kinetic energy within this generalized Davydov--Chaban approach reads 
\begin{equation}
T_{(\beta_2 ,\beta_3)vib} = -\sum_{\lambda=2,3} {\hbar^2 \over 2 B_\lambda}
{1\over \beta_\lambda^3} {\partial \over \partial \beta_\lambda}
\beta_\lambda^3 {\partial \over
\partial \beta_\lambda}.
\end{equation}
Using wave functions of the form
\begin{equation}
\psi^{\pm} (\beta_2 ,
\beta_3)=(\beta_2\beta_3)^{-3/2}\varphi^{\pm} (\beta_2 ,\beta_3),
\end{equation}
which is a straightforward generalization of Eq. (\ref{wfChaban}),
the kinetic energy takes the form 
\begin{equation}
\widetilde{T}_{(\beta_2 ,\beta_3)vib} = -\sum_{\lambda=2,3}\left ( {\hbar^2\over 2 B_{\lambda}}
{\partial^2 \over \partial \beta_{\lambda}^2}
+{3\hbar^2 \over 8B_{\lambda} \beta_{\lambda}^2}\right ) ,
\end{equation}
where again the second term is pushed into the effective potential, as in Eq. (\ref{Tvibeff}). 

From the considerations given above, it becomes clear that the kinetic energy term used in Refs. \cite{AQOA,Dzy,Den},
as well as in the present work, is based on the following assumptions:

1) The $\gamma$ degree of freedom is frozen from the very beginning, thus reducing the degrees of freedom to four
($\beta$, three Euler angles) in the case of pure quadrupole deformation, and to five ($\beta_2$, $\beta_3$, three 
Euler angles) in the case of simultaneous presence of quadrupole and octupole deformations.   

2) Vibration-rotation cross terms are ignored, making the inertial tensor diagonal and allowing the rotated coordinate system 
to be the principal inertial (body) system.

\subsection*{D2. Moments of inertia } 

Using the standard Bohr expression for the nuclear surface in the body-fixed frame, 
given by \cite{Bohr}
\begin{equation}
R(\theta,\phi)= R_0 \left[ 1+\sum_{lm} a_{lm} Y_{lm}(\theta,\phi)\right],
\end{equation} 
where $Y_{lm}(\theta,\phi)$ stands for the spherical harmonics, 
ignoring vibration-rotation cross terms as above,
and assuming that only the even components ($a_{30}$, $a_{3\pm2}$) of the octupole parameters
are non-vanishing,   
we obtain for the moments of inertia in the octupole degree of freedom the expressions \cite{Lipas,Davids}
\begin{eqnarray}
\mathfrak{J}_1^{(3)}&=&B_3(6a_{30}^{2}+2\sqrt{30}a_{30}a_{32}+8a_{32}^{2})\label{J1}\\
\mathfrak{J}_2^{(3)}&=&B_3(6a_{30}^{2}-2\sqrt{30}a_{30}a_{32}+8a_{32}^{2})\label{J2}\\
\mathfrak{J}_3^{(3)}&=&8B_3a_{32}^{2}
\label{J3},
\end{eqnarray}
which in the axial case ($a_{30}=\beta_3$, $a_{32}=0$) give
\begin{eqnarray}
\mathfrak{J}_1^{(3)}=\mathfrak{J}_2^{(3)}=6B_3\beta_3^{2}, \ \ \ \mathfrak{J}_3^{(3)}=0.
\label{J2axi}
\end{eqnarray}
Different expressions for the moments of inertia are obtained if one considers the odd components 
($a_{3\pm1}$, $a_{3\pm3}$) as the non-vanishing ones \cite{Lipas}. Here we make the assumption, as in Ref. 
\cite{Lipas,Davids}, that for low-lying collective negative parity states the even components play the main role, 
since their contributions to the shape are more symmetric, a property usually associated with lower energy 
configurations. 

For the moments of inertia in the quadrupole degree of freedom we use the standard expression \cite{Bohr}
\begin{eqnarray}
\mathfrak{J}_k^{(2)}=4B_2\beta_2^2 \sin^{2}\left (\gamma -\frac{2}{3}\pi k \right ),
\end{eqnarray}
which in the axial case ($\gamma =0$) gives 
\begin{eqnarray}
\mathfrak{J}_1^{(2)}=\mathfrak{J}_2^{(2)}=3B_2\beta_2^{2}, \ \ \ \mathfrak{J}_3^{(2)}=0.
\label{J3axi}
\end{eqnarray}

Collecting (\ref{J2axi}) and (\ref{J3axi}) into the axial quadrupole-octupole
moment of inertia one gets
\begin{eqnarray}
\mathfrak{J}_1^{(2+3)}=\mathfrak{J}_2^{(2+3)}=\mathfrak{J}^{(2+3)}=3B_2\beta_2^{2}+6B_3\beta_3^{2},
\label{J23axi}
\end{eqnarray}
which gives exactly the denominator in the angular momentum part of Hamiltonian (\ref{eq:e1})
\begin{eqnarray}\label{totalInertia}
\frac{1}{2\mathfrak{J}^{(2+3)}}=\frac{1}{6(B_2\beta_2^{2}+2B_3\beta_3^{2})} .
\end{eqnarray} 

From the considerations given above, it becomes clear that the moment of inertia term used in Refs. \cite{AQOA,Den},
as well as in the present work, is based on the following assumptions:

1) Vibration-rotation cross terms are ignored, as in the case of the kinetic energy.

2) Only the axial components of deformation are taken into account, both in the quadrupole and in the octupole degree of freedom,
based on the qualitative expectation that more symmetric configurations would lie lower in energy.  

It should be noted that in Ref. \cite{Dzy} an expression $3(B_2\beta_2^{2}+B_3\beta_3^{2})$ has been used for the moment of inertia. 

\subsection*{D3. Separation of variables}

Exact separation of the $\beta$ and $\gamma$ variables in the framework of the Bohr Hamiltonian 
can be achieved by considering potentials of the form $u(\beta,\gamma)=v(\beta)+ w(\gamma)/\beta^2$
\cite{Wilets,Fortunato}. In contrast, when the potential is of the form $u(\beta,\gamma)= v(\beta) + w(\gamma)$, 
only approximate adiabatic separation of variables can be tried, as in the case of the X(5) critical point symmetry 
\cite{IacX5,Bijker}. In the case of X(5), a $\beta^2$ term survives in the differential equation involving the $\gamma$ variable,
replaced in the adiabatic approximation by the average value $\langle \beta^2 \rangle$. The accuracy of this approximation 
has been tested in Ref. \cite{Caprio72} and the limits of its validity have been pointed out. The recently developed Algebraic 
Collective Model \cite{Rowe735,Rowe753,Caprio672,Welsh} offers a path for avoiding this approximation by performing rapidly converging exact numerical calculations instead of pursuing approximate analytical solutions.  

In the present case, the $\gamma$ variable has been ``frozen'' from the very beginning, following the Davydov--Chaban approach
\cite{Chaban}, as explained in Appendix D1. Therefore, no question of separating the $\beta$ and $\gamma$ variables appears. However,
separation of the $\beta$ and $\phi$ variables is desirable, in order to achieve analytical solutions in closed form. 
By analogy to the X(5) situation described above, a potential of the form  $v(\tilde \beta,\phi) = u(\tilde\beta) + w(\tilde \phi^\pm)$
has been chosen and adiabatic separation of variables has been tried, taking advantage of the fact that the $w(\tilde \phi^\pm )$ potential
is supposed to be of the form of two very steep harmonic oscillators centered at the values $\pm \phi_0$. Because of the steepness
of the oscillators it is plausible to use the adiabatic approximation in the differential equation involving the $\tilde \beta$ variable 
[Eq. (\ref{eq:e9})], by replacing the variable $\phi$ by $\pm\phi_0$. Again in analogy to the X(5) case mentioned above, 
a $\langle \tilde\beta^2 \rangle$ term remains in the differential equation involving the $\tilde\phi$ degree of freedom [Eq. (\ref{eq:e10})].
There is no need to explicitly determine $\langle \tilde\beta^2 \rangle$, since it enters the parameter $b$ [Eq. (\ref{eq:e22})], determined 
from $E2$ transitions as described in subsection III.B.1.
 
In other words, we exploit for the separation of variables the fact that the $w(\tilde \phi^\pm )$ potential
is supposed to be of the form of two very steep harmonic oscillators centered at the values $\pm \phi_0$.
This makes the adiabatic approximation of $\phi$ by $\pm \phi_0$ plausible, isolating the two very steep harmonic oscillators 
in the $\phi$ equation and leaving the rest of the terms in the $\tilde \beta$ equation. 
An alternative possibility is to consider a potential of the form $v(\tilde \beta,\phi) = u(\tilde\beta) + w(\tilde \phi^\pm)/{\tilde\beta^2}$.
Then the separation of variables will become exact, but the distribution of terms in the two equations will be different.

The adiabatic approximation used here, based on two very steep harmonic oscillators, does have a cost. It is well known 
that the correct description of the parity splitting, usually depicted as the odd-even staggering of the energy levels
of the ground state band and the negative parity band, requires a finite barrier between the two wells, which is angular momentum dependent \cite{Jolos49,Jolos587,Jolos60}. In the present approach a practically infinite barrier between the two 
wells is used for all angular momenta. This has as a consequence that the theoretical predictions for the low lying negative parity states (especially for $1^-$ and $3^-$) are poor, as pointed out in subsection III.A.  

It should be mentioned that the Bohr Hamiltonian has been solved for the potential $1/\sin ^2(3\gamma)$ 
[resembling the last fractional term in Eq. (\ref{eq:e9})], possessing a minimum at $\gamma=\pi/6$, first by replacing
the $\gamma$ variable in the moments of inertia by its expectation value, $\gamma_0$, and subsequently avoiding this approximation
\cite{DeBaerde}, the results revealing the approximation to be a good one. Future tests of similar nature in the present framework are desirable.

\subsection*{D4. Comparison to other approaches}

As it has already been mentioned in the Introduction, a more general approach has been developed by
Bizzeti and Bizzeti-Sona \cite{Bizzeti70,Bizzeti77}, in which nonaxial 
contributions, small but not frozen to zero, are taken into account. It is worth commenting briefly
on the relation between the two approaches.

1) In the AQOA approach, no nonaxial contributions are taken into account. As a result,
all variables related to nonaxiality in Ref. \cite{Bizzeti70} are vanishing, the matrix of inertia 
(Table I and II in Ref. \cite{Bizzeti70}) becoming diagonal. 

2) Because of the same reason, in the invariants up to fourth order reported in Table VIII of Ref. \cite{Bizzeti70},
only the first term in each invariant, containing only $\beta_2$ and/or $\beta_3$, is surviving. 
In the present approach only the invariants up to second order, being equal to $\beta_2^2$ and $\beta_3^2$, are used. 

3) In Ref. \cite{Bizzeti70}, in addition to the infinite square well potential, a harmonic oscillator potential
proportional to the square of $x=\sqrt{2} \beta_3/\beta_2$ has been used. In the present approach,
$\beta_3/\beta_2 = \tan\phi \sqrt{B_2/B_3}$. Therefore the two potentials coincide, up to constant factors, 
for small angles, for which $\tan\phi \approx \sin\phi \approx \phi$. 

4) The total moment of inertia appearing in Ref. \cite{Bizzeti70} [Eq. (32)], coincides with the total moment
of inertia used here [Eq. (\ref{totalInertia})], if the nonaxial variables vanish, as seen from Eq. (6a) of Ref. \cite{Bizzeti70}.

\end{document}